\documentclass[prd,aps,nofootinbib,floatfix,12pt,longbibliography]{revtex4}
\usepackage{changes}
\usepackage{amsfonts,amsmath,amssymb}
\usepackage{graphicx,color,epsfig,epstopdf}

\newcommand{\nn}{\nonumber}
\newcommand{\be}{\begin{equation}}
\newcommand{\ee}{\end{equation}}
\newcommand{\bea}{\begin{eqnarray}}
\newcommand{\eea}{\end{eqnarray}}
\newcommand{\ba}{\begin{array}}
\newcommand{\ea}{\end{array}}
\newcommand{\bi}{\begin{itemize}}
\newcommand{\ei}{\end{itemize}}

\newcommand{\mce}{{\mathcal E}}
\newcommand{\mcf}{{\mathcal F}}
\newcommand{\mch}{{\mathcal H}}

\newcommand{\difd}{\mathrm d}

\renewcommand{\vec}[1]{\mbox{\boldmath $#1 \!\!$ \unboldmath}}

\newcommand{\nslash}{\kern 0.2 em n\kern -0.50em /}
\newcommand{\kslash}{\kern 0.2 em k\kern -0.45em /}
\newcommand{\qslash}{\kern 0.2 em q\kern -0.45em /}
\newcommand{\pslash}{\kern 0.2 em p\kern -0.50em /}
\newcommand{\rslash}{\kern 0.2 em r\kern -0.50em /}
\newcommand{\sslash}{\kern 0.2 em s\kern -0.50em /}
\newcommand{\Sslash}{\kern 0.2 em S\kern -0.50em /}
\newcommand{\Pslash}{\kern 0.2 em P\kern -0.50em /}
\newcommand{\Dslash}{\kern 0.2 em D\kern -0.65em /\kern 0.15em}
\newcommand{\lf}{\left}
\newcommand{\rg}{\right}

\newcommand{\xbj}{x_B}                   

\begin{document}
\title{Accessing {Compton Form Factors} at the Electron Ion Collider in China: An Impact study on $\textbf{Im} \mce$}
\author{Xu Cao \footnote{Email: caoxu@impcas.ac.cn}}
\affiliation{Institute of Modern Physics, Chinese Academy of Sciences, Lanzhou 730000, China}
\affiliation{University of Chinese Academy of Sciences, Beijing 100049, China}
\affiliation{Research Center for Hadron and CSR Physics, Lanzhou University and Institute of Modern Physics of CAS, Lanzhou 730000, China}






\author{Jinlong Zhang \footnote{Email: jlzhang@sdu.edu.cn}}
\affiliation{Key laboratory of particle physics and particle irradiation (MOE), Shandong University, Qingdao 266237, China}





\begin{abstract}
  We estimate the impact of asymmetry measurements of Deeply Virtual Compton Scattering (DVCS) with transversely polarized proton beam taken at a future Electron Ion Collider in China (EicC) on the extraction of Compton Form Factors (CFFs). The CFFs extracted from an analysis based on artificial neural-network approach are reweighted by means of pseudo-data generated in the expected kinematic region of EicC. We find a remarkable improvement in the extraction of CFF $\textmd{Im} \mce$, especially at the range of parton momentum fraction $x \sim$ 0.01, thus hinting for a future experimental probe of the parton orbital angular momentum. This work casts a glance at a practical implementation of Bayesian reweighting method on CFFs' impact study.

\end{abstract}

\date{\today}
\maketitle
\newpage
\section{Introduction} \label{sec:intro}

\begin{figure}
  \begin{center}
  {\includegraphics*[width=0.5\textwidth]{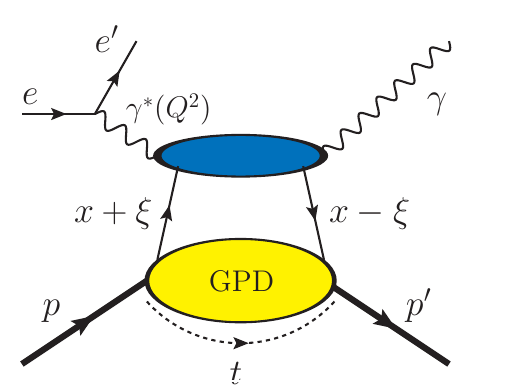}}
    \caption{The handbag diagram of Deeply Virtual Compton Scattering on the proton $e p \to e' p' \gamma$ at leading twist and leading order. Here $Q^2 = -q^2$ is the virtuality of the exchanged photon between the initial electron and proton with $q$ the virtual photon 4-vector.
    The $x$ is the average fractional longitudinal momentum of the active parton, and $\xi$ is half the difference of longitudinal momentum fractions between the initial and the final parton.
    The Mandelstam variable $t = (p - p')^2$ is the squared four-momentum transfer between the initial and final proton.
    }
    \label{fig:handbag}
  \end{center}
\end{figure}

Generalized parton distributions (GPDs), theorized to image the three dimensional nucleon structure in the joint phase space of transverse spatial position and longitudinal momentum, can be probed by exclusive scattering processes.
The cleanest of them, the Deeply Virtual Compton Scattering (DVCS)~\cite{Mueller:1998fv,Ji:1996nm,Ji:1996ek,Radyushkin:1996nd,Radyushkin:1997ki} is illustrated by a handbag diagram in Fig.~\ref{fig:handbag}.
The DVCS amplitudes in the Bjorken limit can be decomposed either into helicity amplitudes, or, equivalently into complex structure functions, \emph{viz.} Compton Form Factors (CFFs), which are to be measured experimentally~\cite{Ji:1998pc,Diehl:2003ny,Belitsky:2005qn,Guidal:2013rya}.
The CFFs $\mch$, $\mce$, $\widetilde{\mch}$ and $\widetilde{\mce}$ are convolution of corresponding chiral-even GPDs $H$, $E$, $\widetilde{H}$, and $\widetilde{E}$, respectively with renormalized coefficient functions calculable at any order of perturbative QCD (pQCD).
Several global extraction of CFFs from DVCS experimental data are now released within neural network~\cite{Kumericki:2011rz,Cuic:2020iwt,Berthou:2015oaw,Moutarde:2019tqa,Grigsby:2020auv}.
Retrieving GPDs from CFFs, known as the deconvolution problem, is still a challenge endeavour~\cite{Kumericki:2007sa,Kumericki:2009uq,Moutarde:2018kwr,Kriesten:2021sqc,Bertone:2021yyz} and steps forward only recently to neural network modelling~\cite{Dutrieux:2021wll}.
Through the gravitational form factors (GFFs), GPDs can also decipher the mechanical properties, e.g. mechanical radius, pressure and shear force distributions of internal nucleon~\cite{Polyakov:2002yz,Goeke:2007fp,Lorce:2018egm,Polyakov:2018zvc}, though suffered from big systematic errors at present~\cite{Shanahan:2018nnv,Burkert:2018bqq,Kumericki:2019ddg}.

The complete and precise extraction of all GPDs from data of exclusive processes puts the highest demands on both theory and facility.
The cutting-edge experiment of GPDs have been ongoing during the passed two decades by ZEUS~\cite{Chekanov:2003ya,Chekanov:2008vy}, H1~\cite{Adloff:2001cn,Aktas:2005ty,H1:2009wnw,Aaron:2009ac}, HERMES~\cite{Airapetian:2012pg}, COMPASS~\cite{COMPASS:2018pup}, Hall A~\cite{Defurne:2017paw,Defurne:2015kxq} and CLAS~\cite{dHose:2016mda} and are proposed to continue at JLab update~\cite{Hyde:2011ke,Burkert:2021rxz,Fucini:2021psq,Afanasev:2021twk} and also future electron-ion colliders~\cite{Deshpande:2005wd,Accardi:2012qut,AbdulKhalek:2021gbh,CAO:2020EicC,CAO:2020Sci,Anderle:2021wcy}.
Available data, limited in the valence region, give much more constraint on GPD $H$ than the other GPDs, as recognized by VGG~\cite{Vanderhaeghen:1999xj,Dupre:2016mai} and Goloskokov-Kroll (GK) 
models~\cite{Goloskokov:2005sd,Goloskokov:2006hr,Goloskokov:2007nt,Goloskokov:2008ib,Goloskokov:2009ia,Goloskokov:2011rd,Goloskokov:2013mba,Goloskokov:2014ika}.
GPD $E$ appears
in measurements with {neutron or} transversely polarized nucleon, which is pioneering, however, suffering from scarcity and sizable statistical uncertainties~\cite{Ye:2006pe,Murray:2007zzb,Airapetian:2008aa,Airapetian:2011uq}.
{The GPD E, has for partner the Sivers function considering that they do not depend on quark helicity and both involve a flip of nucleon helicity.}
The GPD $E$ is of its own importance in the sense of its involvement to the parton orbital angular momenta (OAM) \cite{Ji:1996ek}, which plays an important role in the proton spin decomposition~\cite{Goloskokov:2008ib,Diehl:2013xca,Kroll:2020jat}.
From this perspective, DVCS measurements with a transverse polarized proton beam are proposed as one of the priority programs at the proposed Electron-Ion Collider in China (EicC), designed to collide 3.5 GeV polarized electron beam of 80\% polarization with 20 GeV polarized proton beam of 70\% polarization at instantaneous luminosity of $2 \times 10^{33}$ cm$^{-2}$ s$^{-1}$ or higher~\cite{CAO:2020EicC,CAO:2020Sci,Anderle:2021wcy}.
EicC aims to bridge the kinematic coverage usually referred as sea quark region between JLab and EIC at Brookhaven National Laboratory (BNL).
In view of the required uncertainty at planned facilities in order to make substantial progress in the understanding of the DVCS process, an unbiased determination of CFFs or GPDs from all existing measurements is particularly relevant for future experimental design.
For the time being, a global extraction of GPDs from full data sets has not been accomplished yet, but it does make much progress at the level of CFFs by several groups~\cite{Kumericki:2011rz,Cuic:2020iwt,Berthou:2015oaw,Moutarde:2019tqa,Grigsby:2020auv} and at the cross section level by FemtoNet deep neural network~\cite{Grigsby:2020auv,Almaeen:2022imx}.
Among them, PARTONS framework~\cite{Berthou:2015oaw,Moutarde:2019tqa,Aschenauer:2022aeb} is publicly available in a manner of open source and its unbiased uncertainties propagation embodied into replica method would serve as a remarkable toolkit for our purpose.

In this paper, we introduce Bayesian reweighting strategy to investigate the impact of pseudo-data at EicC on the extraction of CFFs.
Specifically, in Sec. \ref{sec:pseudo-data} we present the methodology applied to generate pseudo-data of transversely polarized proton beam-spin asymmetry within the kinematic region of EicC.
After introducing the reweighting tool used to perform our analysis in Sec. \ref{sec:formula}, we present the sensitivity of CFF $\textmd{Im} \mce$ to specific data samples in a more quantitative manner within PARTONS Artificial neural network (ANN).
We conclude our findings and make further remarks in Sec. \ref{sec:sumry}.

\section{Generation of pseudo-data} \label{sec:pseudo-data}

\subsection{Remarks on theory and kinematics}

At the leading order, DVCS process off proton in Fig.~\ref{fig:handbag} is factorized to the partonic channel $ q \gamma^* \longrightarrow q \gamma $ where an active parton $q$, is re-absorbed into proton that remains intact after collisions.
The observed $e p \to e' p' \gamma$ reaction is the superposition of DVCS with the known Bethe-Heitler (BH) process, {the latter being} the initial and final state radiation described by electromagnetic form factors in quantum electrodynamics (QED).
The GPDs, initially introduced to describe this kind of deeply exclusive processes evolving with the virtuality $Q^2$ of the photon, are dependent on the Mandelstam variable $t$ associated to the four-momentum transfer to the proton, the fractional longitudinal momentum transfer $x$ to the struck parton, and the skewness $\xi$ defining the longitudinal momentum fractions transferred to the parton.
GPDs are widely connected to other interesting physics quantities.
In particular, after performing a Fourier transform with respect to the transverse component of $t$ in the limit case $\xi = 0$, one obtains the impact parameter dependent GPDs~\cite{Burkardt:2000za,Burkardt:2002hr},
which is the probability density to find a parton of longitudinal momentum fraction $x$ with respect to (w.r.t) its transverse distance $\vec{b}_{\bot}$ from the momentum centre of the nucleon,
thus giving tomographic images of the nucleon in 1+2 dimensional joint representation of longitudinal momentum and transverse position.

The first moments of the GPDs $H$, $E$, $\widetilde{H}$, and $\widetilde{E}$ are related to the Dirac, Pauli, axial and pseudoscalar form factors, respectively.
The second moments of the GPDs are relevant to the total angular momenta for quark and gluon through Ji's sum rule~\cite{Ji:1996ek},
\be
J_{q,g}=\frac{1}{2}\int_{-1}^{1}\difd x \,x \lf[H_{q,g}(x,\xi,t =0)+E_{q,g}(x,\xi,t =0) \rg] .
\label{eq:qg-angular-mom}
\ee
Considering that the GPD $H$ (and also $\widetilde{H}$) in the forward direction ($\xi = 0$, $t =0$) reduces to the usual parton distributions constrained by deeply inelastic scattering (DIS) experiments, the GPD $E$ is the genuine interesting issue on the topic of parton angular momentum.
Here $J_{q,g}$combine  gauge-invariantly into the nucleon spin,
\be
\frac{1}{2}=J_q + J_g = \frac{1}{2}\Delta\Sigma + L_q+ J_g \,,
\label{eq:spin-decomp}
\ee
with $\frac{1}{2}\Delta\Sigma$, $L_q$ and $J_g$ being the quark spin angular momentum, quark OAM and gluon total angular momentum respectively.
To study the OAM of the partons, one needs to explore beyond one-dimensional parton distributions.
In view of the quark helicity contribution $\Delta\Sigma$ known also from DIS experiments, the GPD $E$ is the sole missing piece toward the full understanding of quark OAM.
Indeed, GPD $E$ describes the amplitude in terms of the flip of nucleon spin but non-flip of the parton helicities in light-cone frame, implying therefore one unit change of OAM between the initial and final nucleon states.
This provides a practical way to quantifying quark OAM inside the nucleon with solid theory basis.

At leading twist level the GPDs $F$ ($H$, $E$, $\widetilde{H}$, and $\widetilde{E}$), describing the soft structure of the nucleon, enter the cross section of DVCS through its sub-amplitudes, CFFs $\mcf$ ($\mch$, $\mce$, $\widetilde{\mch}$ and $\widetilde{\mce}$), by the convolutions of GPDs over {over variable $x$)}~\cite{Guidal:2013rya,Kumericki:2016ehc},
\be
\mcf (\xi,t,Q^2) = \sum_{q=u,d,s, \cdots} e_q^2
  \int_{-1}^1\difd x\, \left[\frac{1}{\xi - x- i \epsilon} \mp \frac{1}{\xi + x -i \epsilon} \right]
 F^q (x,\xi,t) \,,
\label{eq:CFF}
\ee
where the sum is made over quark flavors $q$ and the upper/lower signs are for the unpolarized GPDs ($H$, $E$) and the polarized GPDs ($\widetilde{H}$, $\widetilde{E}$), respectively.
At leading order we have $\xi \simeq \xbj/(2-\xbj)$.
As genuine observables, CFFs can be precisely measured and separated by various cross sections and asymmetries w.r.t different beam spin, particularly their azimuthal modulations within different kinematic bins ($Q^2$, $\xi$, $t$) (or equivalently $Q^2$, $\xbj$, $t$)~\cite{Diehl:2005pc}.
This makes the first step towards extraction of GPDs globally from worldwide data.

\begin{figure*}
  \begin{center}
  {\includegraphics*[width=0.8\textwidth]{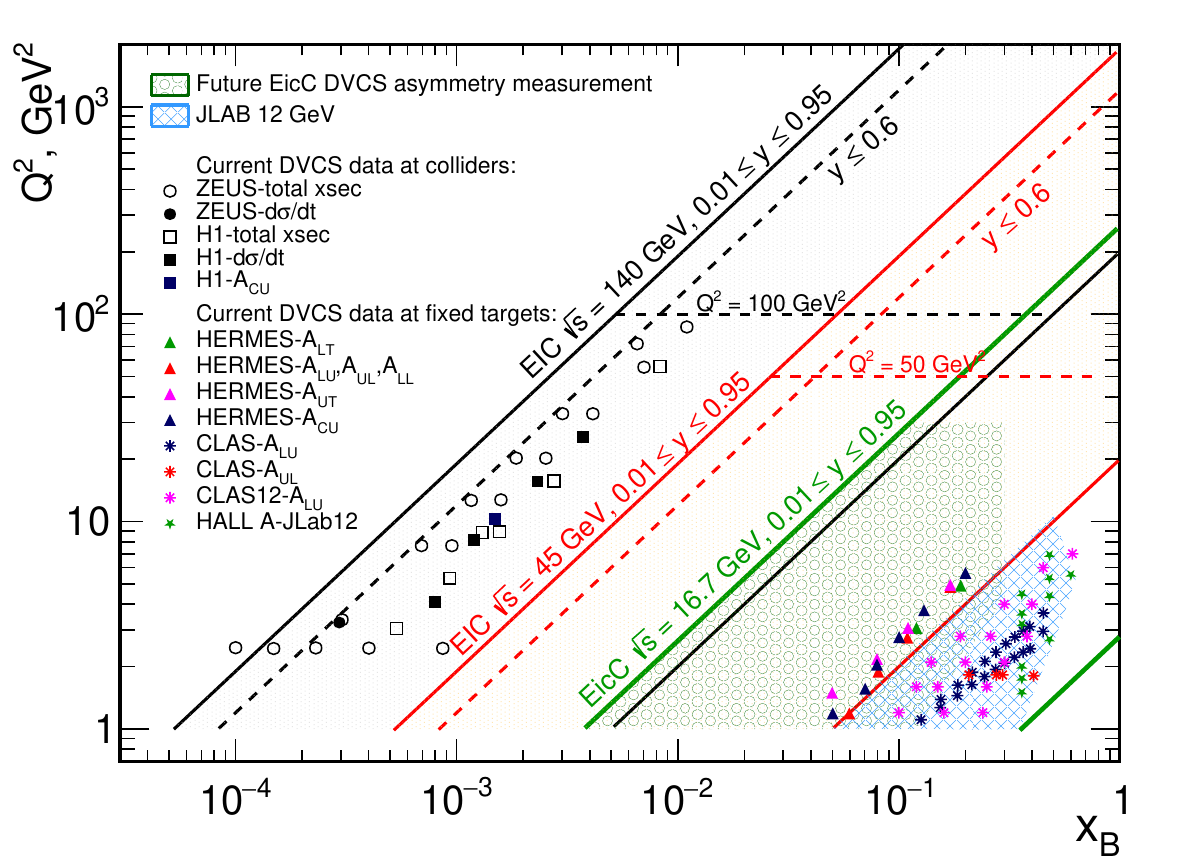}}
    \caption{The kinematic domain of EicC (green hatched area) for measurements of different single spin asymmetries in comparison with the planned JLab 12 GeV program (blue hatched zone) and future EIC at BNL with low (45 GeV) and high (140 GeV) energy scenario. The points are the current DVCS world data at HERA (H1, ZEUS, HERMES) and JLab 6 GeV and 12 GeV (CLAS and Hall A). The figure is adapted from Refs \cite{Accardi:2012qut,Aschenauer:2012ve}.
    }
    \label{fig:kincover}
  \end{center}
\end{figure*}

However, the GPD $E$, and hence CFF $\mce$, involved in the extraction of the quark OAM along the longitudinal axis, surprisingly appears only in {the experiment of neutron target or} a transversely polarized proton beam,
which currently merely available with limited kinematic domain in fix-target experiments at HERMES~\cite{Ye:2006pe,Murray:2007zzb,Airapetian:2008aa,Airapetian:2011uq} and proposed recently by JLab~\cite{dHose:2016mda,Afanasev:2021twk}.
At present only valence region are experimentally accessed with deficient accuracy and expected to be ameliorated by the update of JLab 12 GeV~\cite{Hyde:2011ke}.
It is prerequisite to understanding how sea quarks and gluons behave inside nucleon at future colliders, as depicted in Fig.~\ref{fig:kincover} together with existing measurements.
Nowadays it is widely recognized that such extensive programs with large kinematic coverage and accurate measurements of asymmetries and cross sections are together indispensable for high precision extracting of GPDs.
Under current configuration of the electron and proton beam-energy (3.5 GeV $\times$ 20 GeV), EicC will dramatically extend the experimental study of GPDs with high statistics to the sea quark region of 0.01 $< \xbj <$ 0.1 when restricting to the safely perturbative region 2 $< Q^2 <$ 30 GeV$^2$, and reach down to $\xbj \sim 0.004$ when $Q^2$ approaching to 1 GeV$^2$.
It is feasible to ascertain the uncultivated domain of larger virtuality $Q^2 >$ 30.0 GeV$^2$ with bearable statistics for measuring the differential cross section. It will touch the valence quark region with relatively higher $Q^2$ than that at JLab 12GeV program.
This kinematic coverage of EicC is unique to explore parton spatial tomography of the sea quark inside nucleon, thus complementary to the EIC at RHIC with the aim of understanding glue. EicC is also distinctive in the sense that the interference between the DVCS and the BH processes is more prominent at a lower energy machine, whereas the DVCS process is expected to dominate at high energies.

\subsection{Pseudo-data production}

\begin{figure*}
  \begin{center}
  {\includegraphics*[width=0.8\textwidth]{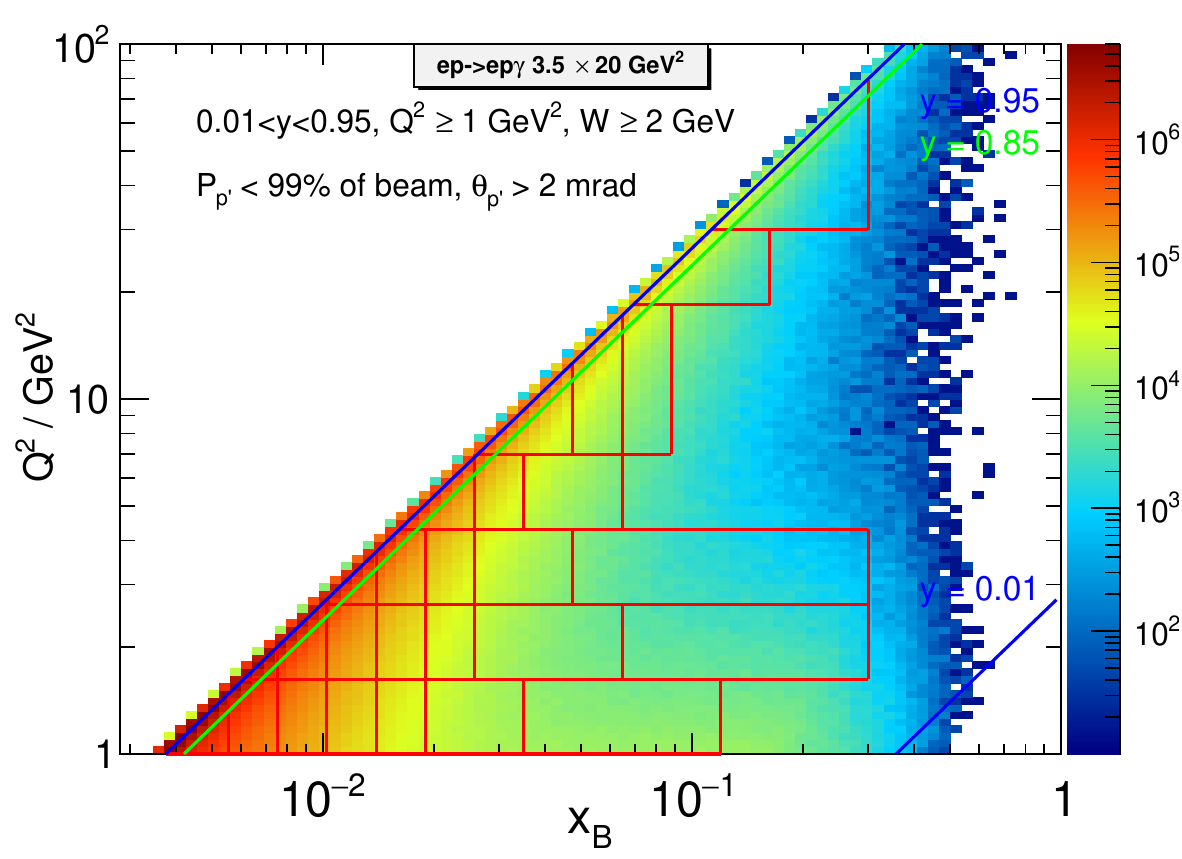}}
    \caption{The events distribution of DVCS in the ($\xbj$, $Q^2$) plane at EicC. The samples contain DVCS, BH, and their interference term. A perfect detector acceptance and efficiency is assumed besides the explicitly shown kinematic cuts (blue lines) and bin scheme (red rectangles). The green $y$ = 0.85 line is only shown for guideline.}
    \label{fig:kinEicC}
  \end{center}
\end{figure*}
\begin{figure*}
  \begin{center}
  {\includegraphics*[width=0.45\textwidth]{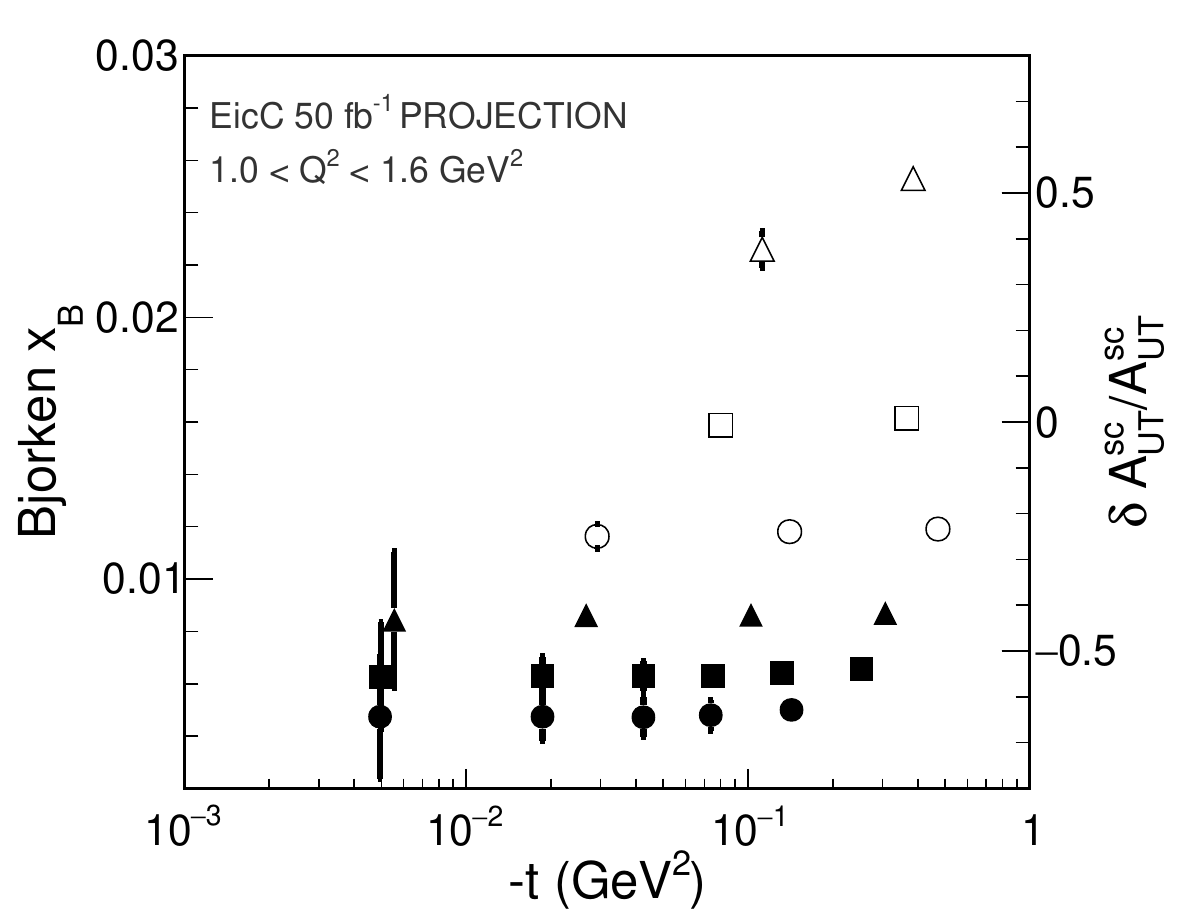}}
  {\includegraphics*[width=0.45\textwidth]{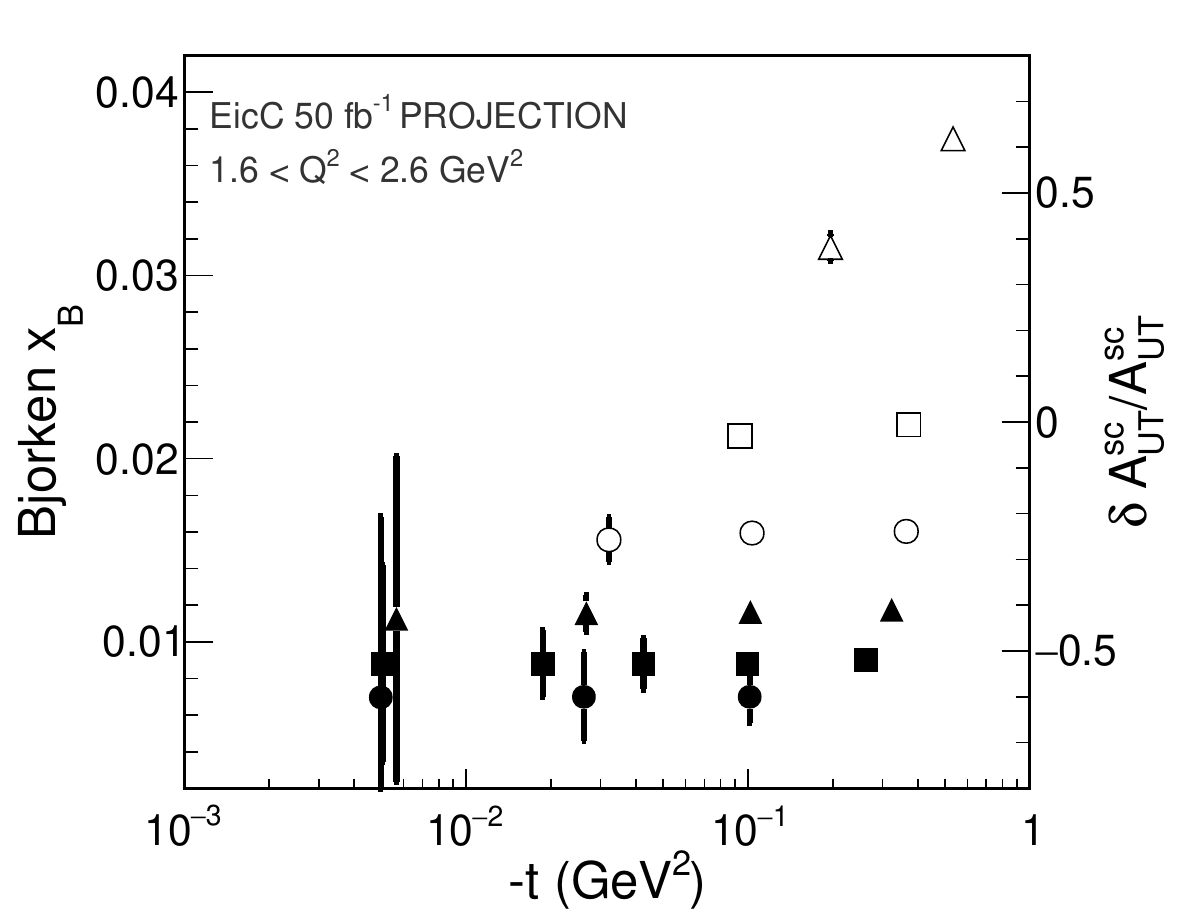}}
  {\includegraphics*[width=0.45\textwidth]{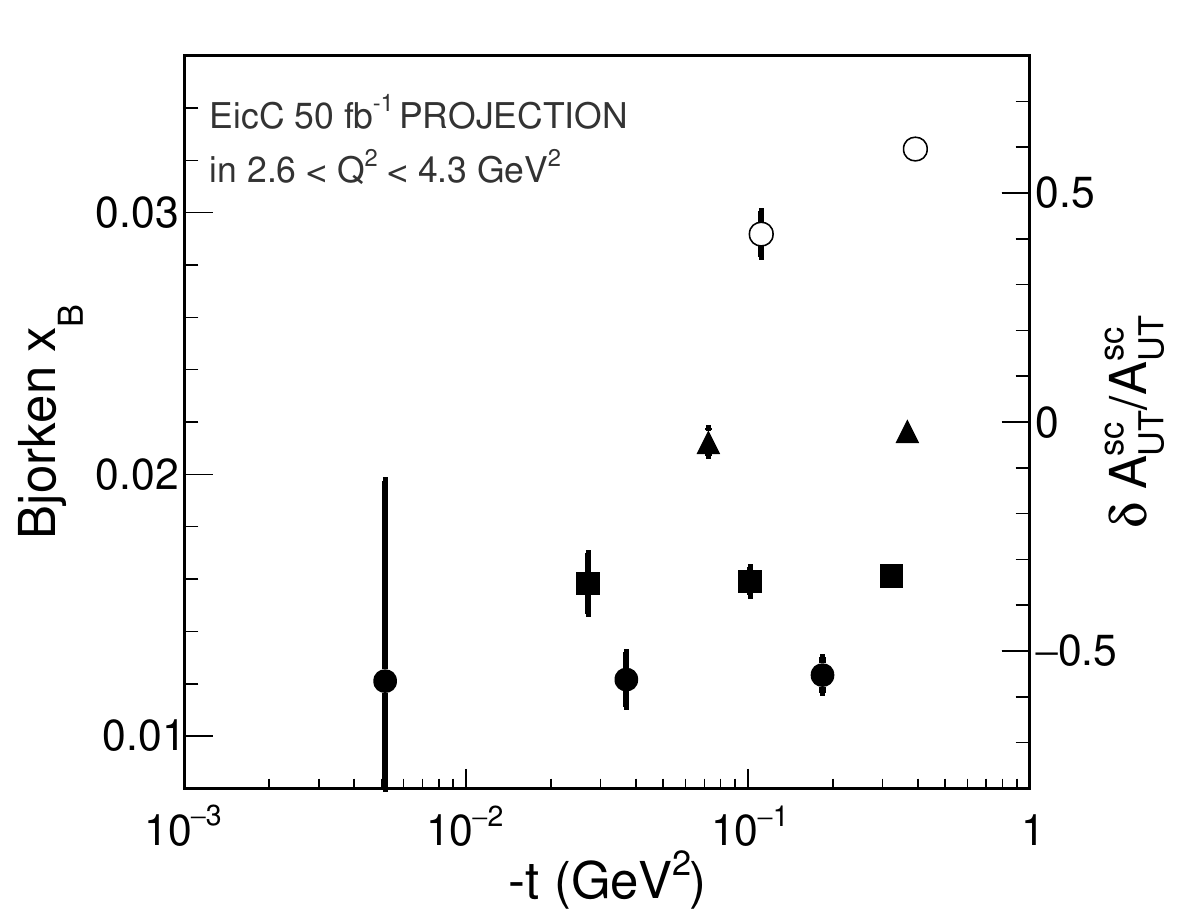}}
  {\includegraphics*[width=0.45\textwidth]{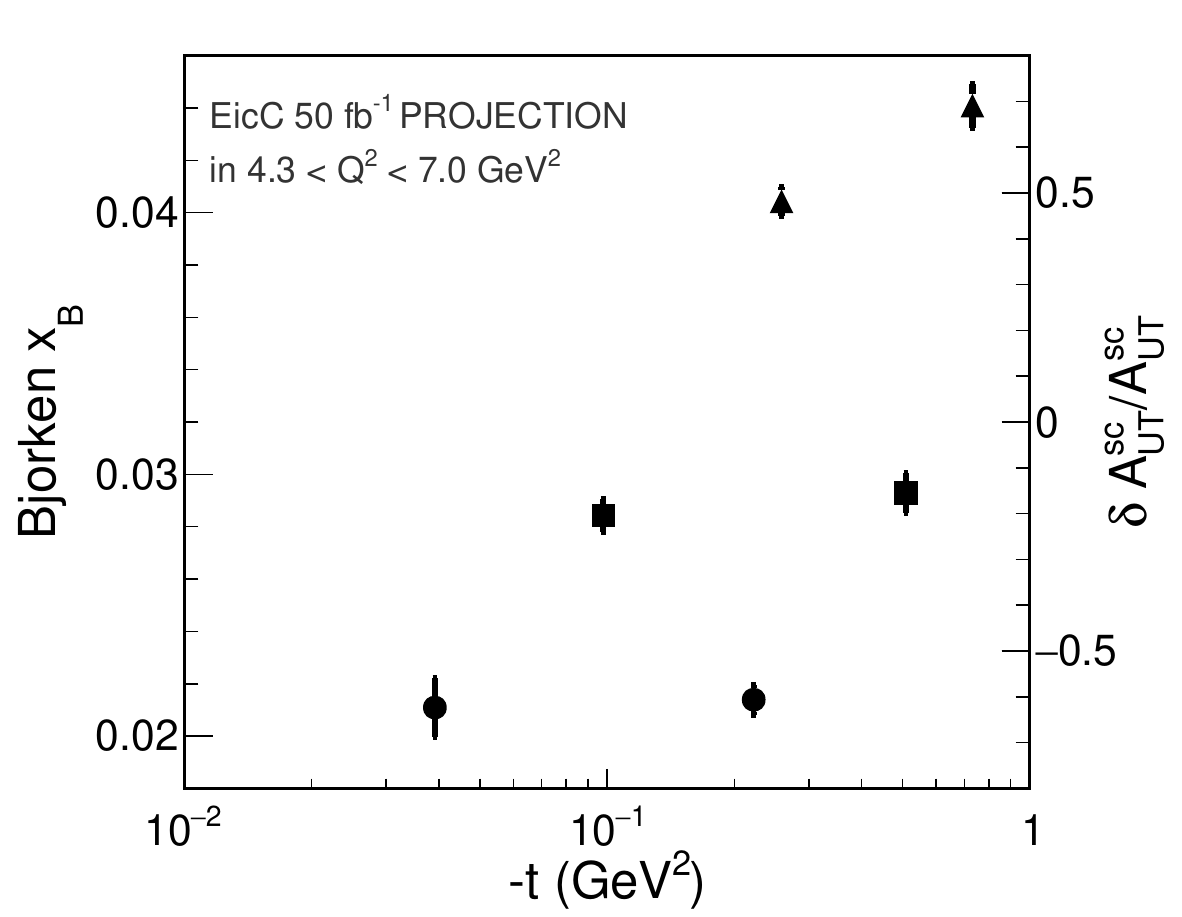}}
  {\includegraphics*[width=0.45\textwidth]{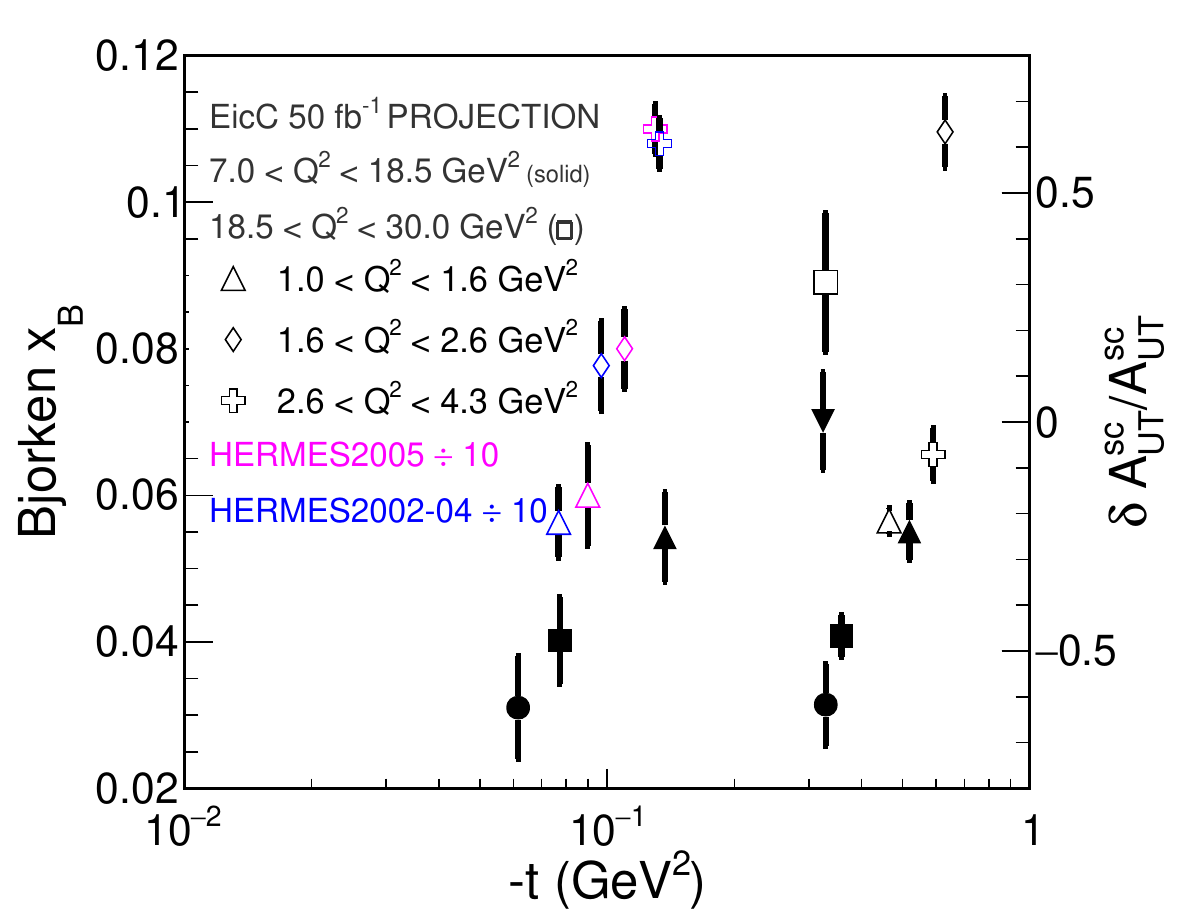}}
    \caption{The projected relative accuracy for $\textrm{A}_{\textrm{UT}}^{\textrm{SC}} \equiv  \textrm{A}_{\textrm{UT}}^{\sin(\phi-\phi_s)\cos\phi}$ asymmetry in the process of DVCS with transversely polarized proton beam in the region 1.0 GeV$^2<Q^2<30.0$ GeV$^2$ under the integrated luminosity 50.0 fb$^{-1}$ at EicC. Only statistics uncertainty is included by MILOU generator.
    The relative uncertainty of each data point should be interpreted using the scale indicated on the right-side vertical axis of the plots.
    The size of $\textrm{A}_{\textrm{UT}}$ is estimated with GK model~\cite{Goloskokov:2009ia,Goloskokov:2011rd,Kroll:2012sm}.
    The black star is the HERMES data of $\textrm{A}_{\textrm{UT,I}}^{\sin(\phi-\phi_s)\cos\phi}$ asymmetry~\cite{Airapetian:2008aa}. The values of $|t|$ bins under the same $Q^2$ are not shown here for simplicity.
    \label{fig:relAUT}}
  \end{center}
\end{figure*}

In the present exploration an integrated luminosity of 50 fb$^{-1}$ is assumed for the generated data samples, corresponding to around 290 days of data taking under the current luminosity design of EicC with a 100\% operational efficiency.
The DVCS and BH processes together with their interference term have been simulated by the Monte Carlo (MC) generator MILOU~\cite{Perez:2004ig}, slightly modified from its original version~\cite{Aschenauer:2013hhw,Aschenauer:2017jsk}.
The DVCS amplitude is evaluated by CFFs tables generated in a GPD-inspired framework to next-to-leading (NLO) and twist-two accuracy incorporating as well NLO GPD evolution~\cite{Belitsky:1999hf}. The exponential $t$-dependence of the DVCS amplitude is introduced with a constant $t$-slope parameter 5.45 GeV$^{-2}$ for the  MILOU steering card.
Fig.~\ref{fig:kinEicC} demonstrates the event samples obtained from MILOU package using the following loose selection criteria:
the invariant mass of $\gamma p'$ in final states are set up  $W>$ 2.0 GeV to isolate the resonance contribution;
the detector acceptance is 0.01 $< y <$ 0.95; the momentum \textrm{P}$_{p'}$ of the final proton in the laboratory frame is smaller than 99\% of the initial proton momentum, and the scattering angle ${\theta}_{p'}$ is bigger than 2 mrad.
The scattered proton acceptance is constrained to be 1.0 $> |t| >$ 0.01 GeV$^2$ with a resolution of $\Delta |t| >$ 0.02 GeV$^2$.
Possibility of 0.002 $< |t| <$ 0.01 GeV$^2$ is scrutinized, critically depending on the final performance of detector.
For comparison, the values of $|t| >$ 0.03 GeV$^2$ and $\Delta |t| >$ 0.03 GeV$^2$ are adopted at EIC simulation~\cite{Aschenauer:2013hhw}.

The ($\xbj$, $Q^2$) bin scheme of Fig. 3 is logarithmically spanning in seven $Q^2$-bins:
\be \nn
[1.0, 1.6],\quad [1.6, 2.6],\quad [2.6, 4.3],\quad [4.3, 7.0],\quad [7.0, 18.5],\quad [18.5, 30.0],\quad [30.0, 80.0]\, \textrm{GeV}^2 \,,
\ee
each of which contains at most 7 $\xbj$-bins. In total 27 kinematic bins in ($\xbj$, $Q^2$) plane are accessible with considerable statistical significance in range of 1.0 GeV$^2 < Q^2 <$ 30.0 GeV$^2$.
Besides, 1 bin at $Q^2 \in$ [30.0, 80.0] GeV$^2$, under good control of statistical uncertainty only for measuring differential cross sections, extends the reachable phase space of EicC to very high $Q^2$.
Each of ($\xbj$, $Q^2$) bins is further divided into $t$-bins, among which 22 have at least 2 $t$-bins, providing the handle of the impact parameter space after Fourier transform.
Altogether 69 bins in ($\xbj$, $Q^2$, $-t$) plane can be individually decomposed into 18 $\phi$-bins $\in$ [0, 2$\pi$], where $\phi$ is the angle between the leptonic plane and the real photon production plane.
The number of events in each kinematic bin is proportional to the $\difd\sigma(\phi,\phi_S)$, the abbreviation of five-fold differential cross section,
\be
\difd\sigma(\phi,\phi_S) \equiv \frac{\difd\sigma^{e p \to e' p' \gamma}}{\difd \xbj \difd Q^2 \difd |t| \difd\phi \difd\phi_S} \,,
\ee
which is the coherent sum of DVCS and BH amplitudes,
\be
\difd\sigma = \difd\sigma_{\textrm{UU}}^{\textrm{BH}} + e_l \difd\sigma_{\textrm{UU}}^{\textrm{I}} + \difd\sigma_{\textrm{UU}}^{\textrm{DVCS}}+ S_\textrm{T} ( e_l \difd\sigma_{\textrm{UT}}^{\textrm{I}} + \difd\sigma_{\textrm{UT}}^{\textrm{DVCS}} ) \,,
\ee
{with the $e_l$ being the electron beam charge in units of the elementary charge.} The $\phi_S$ is the angle between lepton scattering plane and $S_\textrm{T}$, the transverse component of the incoming proton spin
polarization vector that is orthogonal to photon direction.
The transversely polarized proton beam-spin asymmetry $\textrm{A}_{\textrm{UT}}$ is selected as a trial observable, defined in term of the charge-normalized cross sections for opposite orientations of the transverse spin of the nucleon,
\be \label{eq:asymphi}
\textrm{A}_{\textrm{UT}} (x,Q^2) = \frac{\difd\sigma(\phi,\phi_S)-\difd\sigma(\phi,\phi_S+\pi)}{\difd\sigma(\phi,\phi_S)+\difd\sigma(\phi,\phi_S+\pi)} \,,
\ee
which is approximately given by a $\sin{(\phi-\phi_s)}\cos{\phi}$ dependence plus a $\cos{(\phi-\phi_s)}\sin{\phi}$ modulation~\cite{Belitsky:2001ns}.
Under the assumption of dominance of BH term in above denominator it still obtains more or less direct linear dependence on CFFs of $\textrm{A}_{\textrm{UT}}$.
For instance, CFF $\mathcal{E}$ of proton becomes manifest in the $\sin{(\phi-\phi_s)}\cos{\phi}$ module, whose interference part of amplitudes are given by,
\bea
\textrm{A}_{\textrm{UT,I}}^{\sin{(\phi-\phi_s)}\cos{\phi}} &\propto& \mbox{Im}\,
              \Big[-\frac{t}{4M^2}\big({ F_2\mathcal{H}}-{F_1\mathcal{E}}\big)
     +\xi^2\big(F_1+\frac{t}{4M^2}F_2\big)\big(\mathcal{H}+\mathcal{E}\big)\nn\\
   &&\qquad  -\xi^2\big(F_1+F_2)\big(\widetilde{\mathcal{H}}
              + \frac{t}{4M^2}\widetilde{\mathcal{E}}\big)\Big] \,,
\eea
where $\textmd{Im} \mce$ is imaginary part of CFF $\mce$, and $F_{1,2}$ are the nucleon Dirac and Pauli form factor.
The $\cos{(\phi-\phi_s)}\sin{\phi}$ module is complicated by contributions from both CFF $\widetilde{\mathcal{H}}$ and $\widetilde{\mathcal{E}}$,
\bea
\textrm{A}_{\textrm{UT,I}}^{\cos{(\phi-\phi_s)}\sin{\phi}} &\propto& \mbox{Im}\, \big({ F_2\widetilde{\mathcal{H}}}-{F_1 \xi \widetilde{\mathcal{E}}}\big)\,,
\eea
So the $\sin{(\phi-\phi_s)}\cos{\phi}$ module, providing a rare access to the $\textmd{Im} \mce$ with no kinematic suppression of its contribution relative to those of the other CFFs, is what we are really care about for the present study.
{ The full analytic formulas, which relate CFFs with observables at the twist-two level and
include the power-suppressed contributions, are explicitly listed in Eqs. (71, 75) of Ref. \cite{Belitsky:2001ns}.}

The uncorrelated statistical uncertainties in each bin are calculated with the help of likelihood method,
\bea \label{eq:module}
\delta\textrm{A}_{\textrm{UT}}^{\sin{(\phi-\phi_s)}\cos{\phi}} = \frac{\sqrt{2}}{f P_\textrm{T}} \sqrt{ \frac{1 - \langle\textrm{A}_{\textrm{UT}}\rangle}{N_{events}} } \,,
\eea
where $N_{events}$ is the total number of BH/DVCS events obtained after scaling the generated cross sections to integrated luminosity of EicC.
{The $P_T = 70$\% is the transverse polarization of nucleon beam.}
As displayed in Fig.~\ref{fig:relAUT},
the EicC measurements of the $\textrm{A}_{\textrm{UT}}$ with a single angular 
modulation ${\sin(\phi-\phi_s)\cos\phi}$ have rather small statistical uncertainties for a wide kinematic region, as low as a few percent at values of $|t| >$ 0.01 GeV$^2$ judged by the GK model.
This implies that the measurement is actually limited by systematic uncertainties, of a few percentage depending on the facility design, which can be easily incorporated with quadrature addition.
The events are largely accumulated in 0.002 $< |t| <$ 0.01 GeV$^2$ region, giving rise to well-below 0.01 absolute uncertainties.
The relative uncertainties in this region are as high as tens of percent, solely driven by the tiny magnitudes of asymmetries.

\section{Description of the impact study} \label{sec:formula}

\begin{figure*}
  \begin{center}
  {\includegraphics*[width=0.45\textwidth]{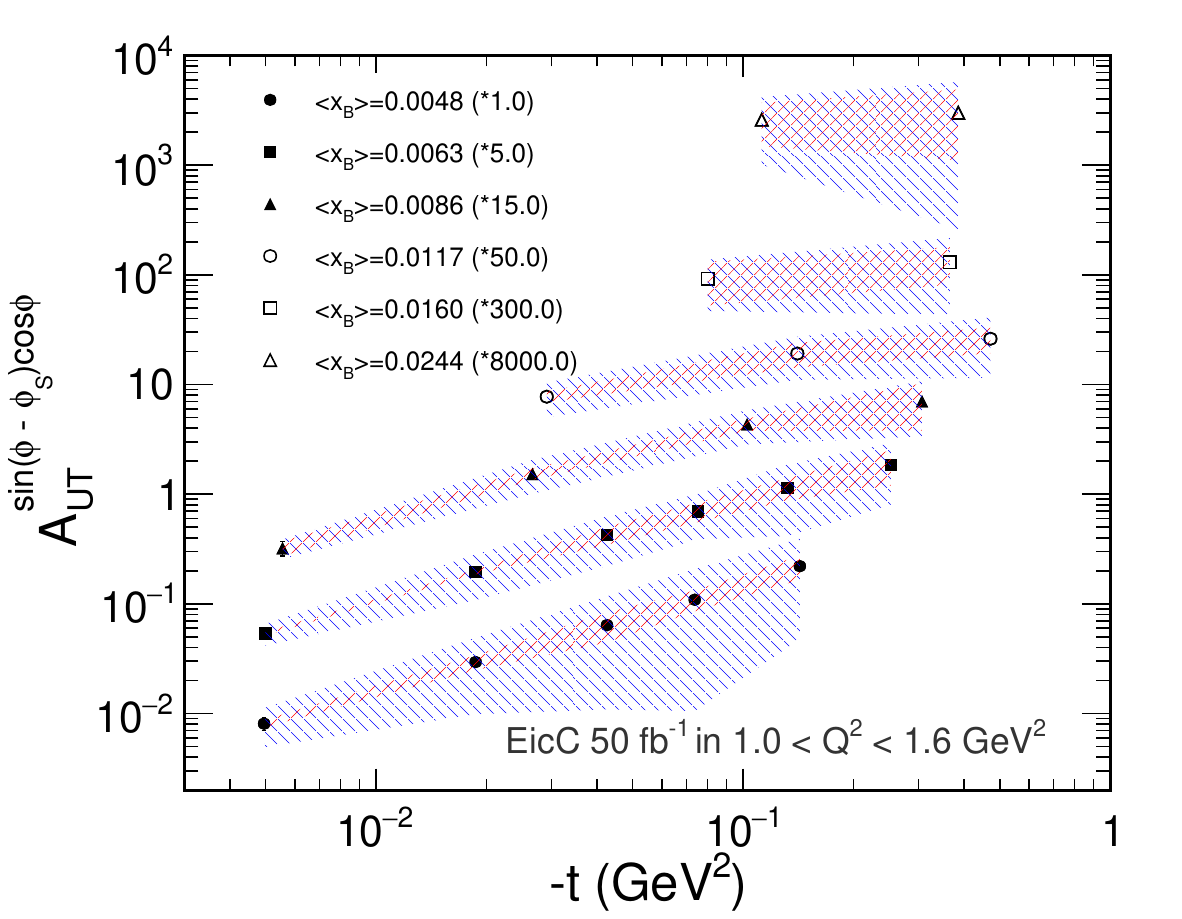}}
  {\includegraphics*[width=0.45\textwidth]{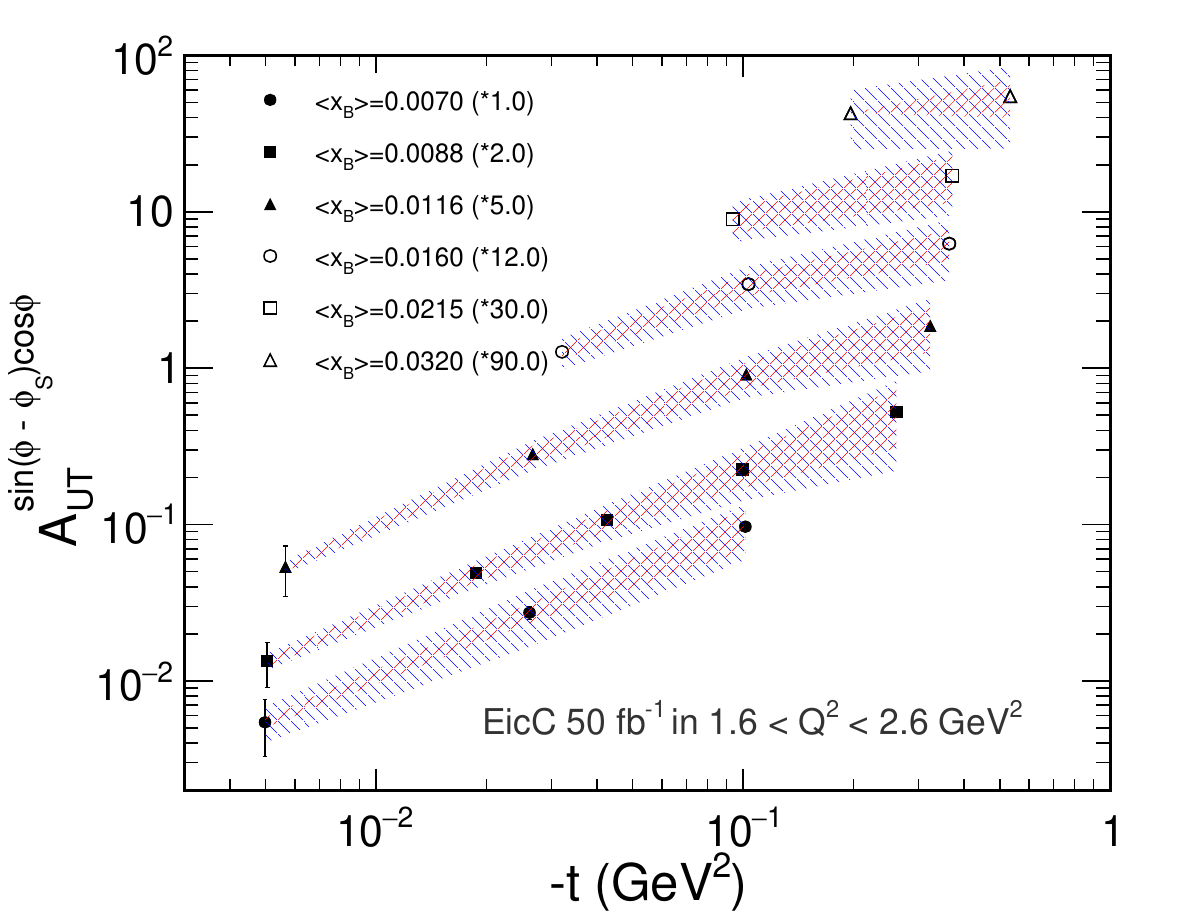}}
  {\includegraphics*[width=0.45\textwidth]{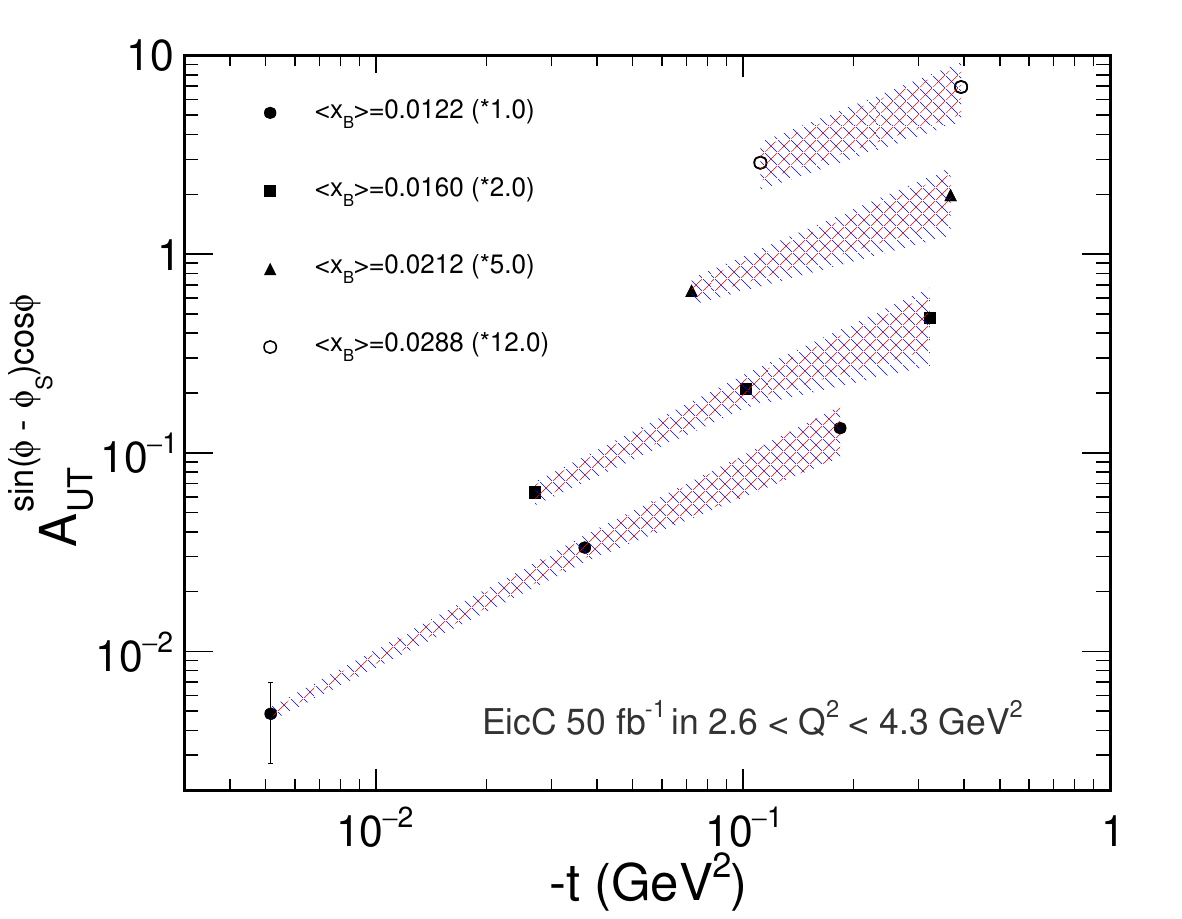}}
  {\includegraphics*[width=0.45\textwidth]{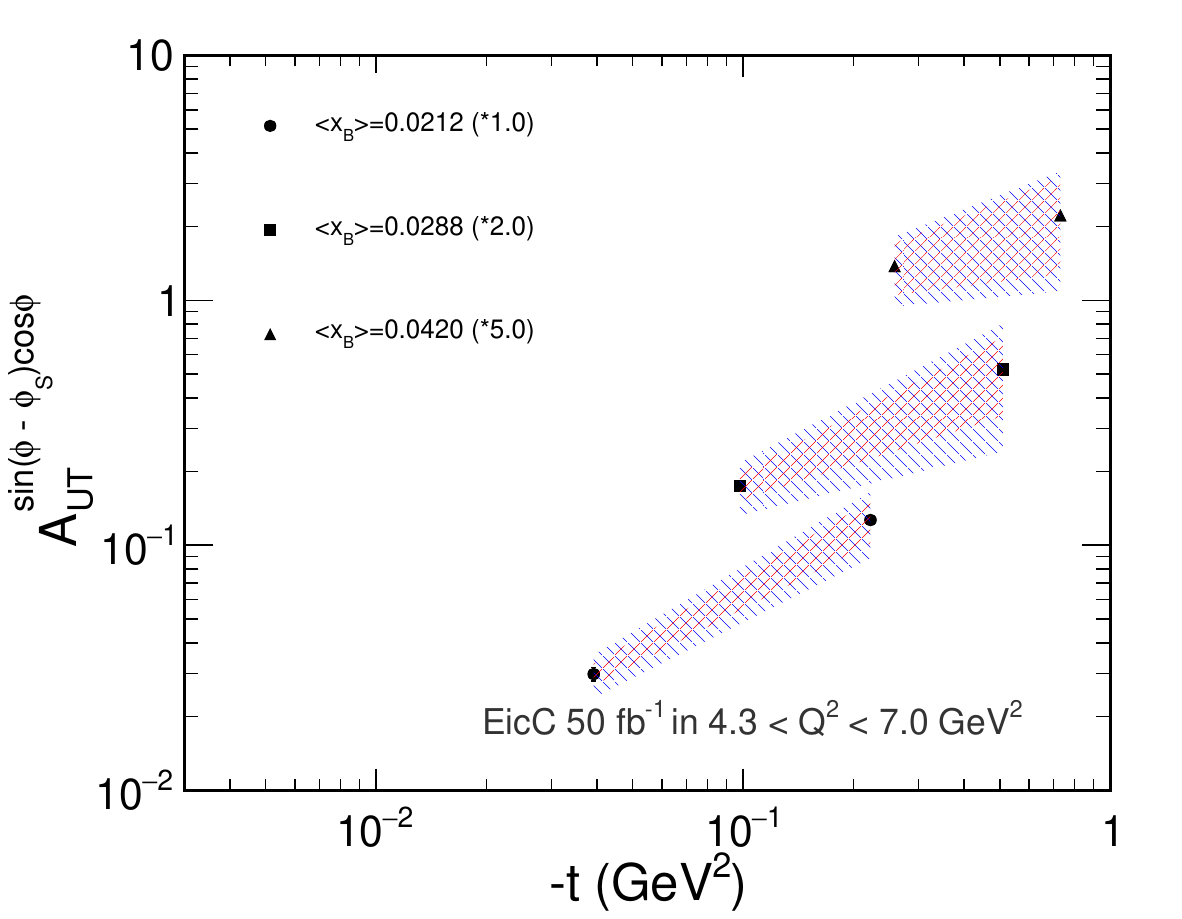}}
  {\includegraphics*[width=0.45\textwidth]{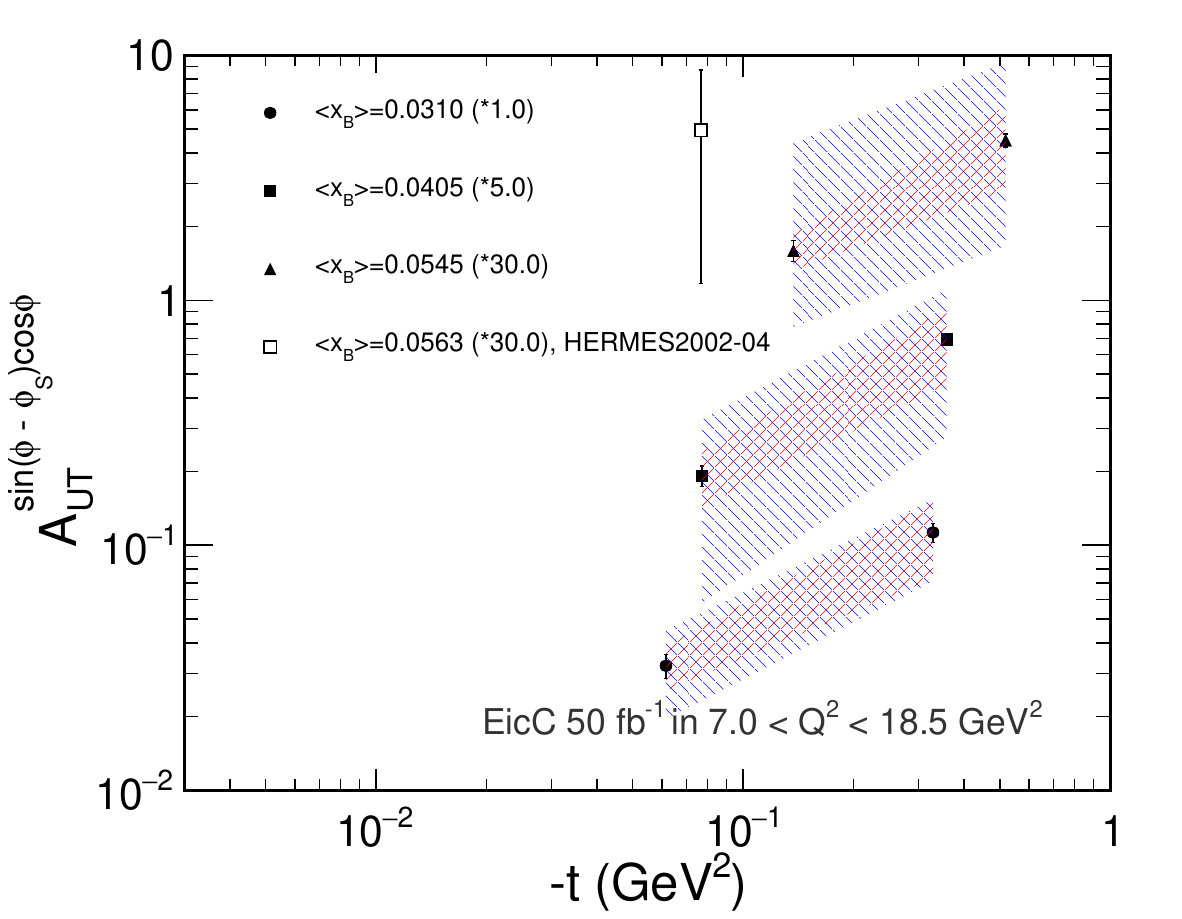}}
    \caption{The pseudo-data of $\textrm{A}_{\textrm{UT}}^{\sin{(\phi-\phi_s)}\cos{\phi}}$ asymmetry generated within the kinematic coverage at EicC.
    The error bars of points correspond to the integrated luminosity 50.0 fb$^{-1}$.
    The blue bands are current constraint in sea-quark region evaluated by PARTONS neural network,
    and {the red bands are those after reweighted by pseudo-data of 0.12 fb$^{-1}$}.
    All the central values are taken from GK model for purpose of demonstration only.}
    \label{fig:logAUT}
  \end{center}
\end{figure*}
\begin{figure*}
  \begin{center}
  {\includegraphics*[width=0.8\textwidth]{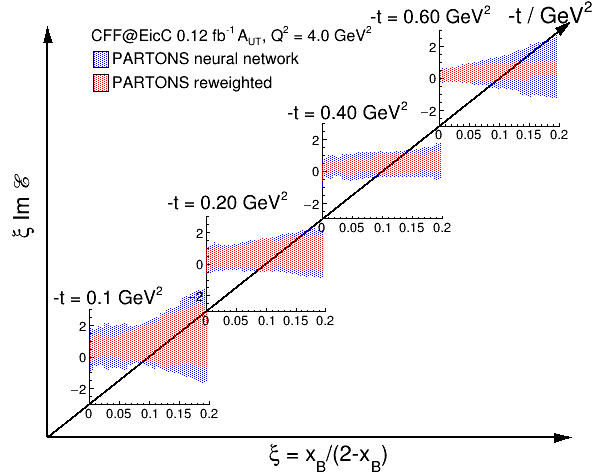}}
    \caption{The $\xi \textmd{Im} \mce$ versus skewness $\xi$ under $Q^2 = $ 4.0 GeV$^2$ and $-t =$ 0.10, 0.20, 0.40, and 0.60 GeV$^2$.
    The blue and red error bands are uncertainties within framework of PARTONS ANN before and after reweighting by the pesudodata of $\textrm{A}_{\textrm{UT}}^{\sin{(\phi-\phi_s)}\cos{\phi}}$ modulation at EicC with the integrated luminosity {0.12} fb$^{-1}$.
    The systematic uncertainty is not considered yet.}
    \label{fig:rewtEicC}
  \end{center}
\end{figure*}

Since ten years ago ANN was utilized to globally extract the CFFs under the assumption of vanishing real part of $\widetilde{\mathcal E}$ and $\widetilde{\mathcal H}$, which are poorly constrained by available data~\cite{Kumericki:2011rz,Kumericki:2019ddg}.
This unbiased method is shown to properly reduce model dependency and propagate the uncertainties in the sense that the obtained CFFs are hardly constrained in kinematic region where data are scarce or deficient.
An open source PARTONS framework was developed and publicly available~\cite{Berthou:2015oaw,Moutarde:2019tqa}, which abandoned the aforementioned assumption and
{extended substantially the set of data used in phenomenological studies.}.
It is an ideal starting point for follow-up Bayesian reweighting technique according to the rules of statistical inference, as proved by NNPDF~\cite{Ball:2010gb,Ball:2011gg}.
To enable this study, we utilize the ensemble of $N_{\textrm{rep}} =$ 101 CFF replicas generated through importance sampling by PARTONS with the input of DVCS data collected in the past two decades. 
The uncertainties of each CFF are independently described by these initial replicas as a function of ($\xi$, $Q^2$, $-t$).
After locally removing the outliers of those replicas with iteratively the 3$\sigma$ rule~\cite{Dutrieux:2021nlz,Dutrieux:2021ehx} and averaging over this distilled ensemble, 
the mean values of  $\textrm{A}_{\textrm{UT}}^{\sin{(\phi-\phi_s)}\cos{\phi}}$ and the predicted size of CFF in sea-quark region, are nearly zero with large standard deviation, reflecting the little constraint power of data in this region. 

{The superscript "SC" stands for $\sin{(\phi-\phi_s)}\cos{\phi}$, now explicitly labeled in caption of Fig.4 but disable in the main text.}
For demonstration purpose, instead of relying on the central values estimated from ANN replicas, we use those given by GK model, $\textrm{A}_{\textrm{UT}}^{\sin{(\phi-\phi_s)}\cos{\phi}}$, see Fig. \ref{fig:logAUT}.
The current integrated luminosity of $\textrm{A}_{\textrm{UT}}^{\sin{(\phi-\phi_s)}\cos{\phi}}$ in sea-quark region is estimated by PARTONS to be only around 0.01 fb$^{-1}$, for comparison see the blue bands in Fig. \ref{fig:logAUT}.
This figure also reflects the uncertainties for the cross sections and the extracted slope parameter.

The mean values of the asymmetry evaluated from PARTONS replicas are then randomly smeared by the statistical uncertainties at EicC.
The generated pseudo-data in this way, labeled as a $N_{\textrm{data}}$ row vector-$\textbf{y}$, are exploited to update the ensemble of replicas by calculating the weight of each replica~\cite{Ball:2010gb,Ball:2011gg},
\bea
\omega_k &=& \frac{1}{Z} (\chi_k^2)^{\frac{N_{\textrm{data}}-1}{2}} e^{-\frac{\chi_k^2}{2}}
\eea
with the normalization factor $Z$ fixed by $\sum_{k=1}^{N'_{\textrm{rep}}}\omega_k = N'_{\textrm{rep}}$.
Note $N'_{\textrm{rep}} < N_{\textrm{rep}}$ due to the removal of outliers.
The $\chi_k^2$ is the goodness of fit indicator between replica and pseudo-data \cite{Paukkunen:2014zia}
\be
\chi_k^2 = (\textbf{y} - \textbf{y}_k) \sigma^{-1} (\textbf{y} - \textbf{y}_k)^T
\ee
where $\textbf{y}_k$ is the row vector of $k$-th replica generated in ANN at the kinematic bins of pseudo-data,
$\sigma$ is the covariance matrix of pseudo-data $\textbf{y}$ and $^T$ denotes matrix transposition.
The number of effective replicas after reweighting are judged in the same spirit of Shannon entropy:
\be
N_{\textrm{eff}} = e^{-\sum_{k=1}^{N'} \omega_k \log\omega_k}
\ee
The newly measurements at EicC is supposed to be much more precise than contemporary uncertainties inferred from ANN as illustrated in Fig. \ref{fig:logAUT}, resulting into a large reduction of uncertainty of CFF.
This is verified by a trial inspection under 50.0 fb$^{-1}$ luminosity of EicC that little effective replicas survive on reweighting.
In order to keep the reweighting procedure reliable, we reconcile to decrease the luminosity to {0.12} fb$^{-1}$.
The number of left effective replicas still relevant is around 1/3 of the initial replicas, keeping the importance sampling off statistical insignificance.

In kinematic terms the factorization in Fig.~\ref{fig:handbag} is valid when the virtuality $Q^2$ of the photon probe is large but the momentum transfer $-t \ll Q^2$ to the nucleon is small compared to this scale of the probe.
In order to match the kinematic cuts in PARTONS, the additional conditions
\be
Q^2 > 1.5~\mathrm{GeV}^2 \,, \qquad \frac{-t}{Q^2} < 0.2
\ee
are applied to the avoid significant higher-twist corrections~\cite{Moutarde:2013qs,Braun:2014sta,Defurne:2015kxq,Defurne:2017paw,Guo:2021ibg}. Then the lowest $Q^2 $ bin [1.0, 1.6] GeV$^2$ at EicC is not included into the reweighting strategy.
The impact of remnant pseudo-data on the extraction of imaginary part of Compton form factors (CFF) ${\mathcal E}$ is displayed in Fig.~\ref{fig:rewtEicC} under $<Q^2> = $ 4.0 GeV$^2$ and several typical $-t$ values ranging from 0.10 to 0.60 GeV$^2$.
The blue bands represent the accuracy driven by the existing data.
The red bands show the accuracy after including the projected $\textrm{A}_{\textrm{UT}}^{\sin{(\phi-\phi_s)}\cos{\phi}}$ data of EicC under the integrated luminosity of {0.12} fb$^{-1}$.
One can see that the uncertainties for the extraction of the imaginary CFF ${\mathcal E}$ is obviously reduced in the sea quark region especially in the large $-t$ region with only hours running of EicC. The impact beyond sea quark region is also unraveled.
So we have observed the statistical relevance of transverse polarization asymmetry measurements at EicC relying on the global extraction of CFF.
An accurate knowledge of CFF ${\mathcal E}$ would have direct consequences on the first glimpse of the quark OAM inside proton.
The large coverage of kinematic range will diminish the uncertainties appearing in Fourier transformation to impact-parameter-dependence, leading to a precise visualization of transverse position space and illumination of 3-dimensional nucleon structure.

Since the new dataset contain a lot of information on the CFFs, necessitating a reweighting globally or a re-training against all the measured and proposed experimental data in order to excavate the full influence of the machine.
The re-training under the same pseudo-data in Fig.~\ref{fig:logAUT} is attempted within KM neural network~\cite{Kumericki:2011rz,Kumericki:2019ddg,Cuic:2020iwt} in EicC white paper~\cite{Anderle:2021wcy}.
There the uncertainty bands of CFF ${\mathcal H}$ and ${\mathcal E}$ are smaller than those in PARTONS under the hypothesis of vanishing real part of $\widetilde{\mathcal E}$ and $\widetilde{\mathcal H}$. 
Still, we monitored a significant reduction of uncertainties of $\textmd{Im} \mce$, as selectively depicted at $-t = 0.2\,{\rm GeV}^2$.
It also considerably ignites the understanding the real part $\textmd{Re} \mce$ in the range of $\xi < 0.1$ when dispersive relations are enabled.

\section{Discussion and Conclusion} \label{sec:sumry}

The understanding of proton spin and 3D structure in conventional quark model tends to be insufficient with desired accuracy~\cite{An:2019tld,Lu:2010dt}.
Though quark OAM are in connection with some of the {transverse momentum dependent parton distributions}, e.g. the Sivers function~\cite{Lu:2006kt,Bacchetta:2011gx,Lorce:2011kd,Thomas:2008ga} or pretzelosity~\cite{Avakian:2008dz,She:2009jq} in a phenomenological way,
its relation to {GPD} $E$ is what genuinely sustained by solid theoretical foundation. 
A precise extracting GPD $E$ and also $H$ from exclusive processes hopefully leads to a quantification of quark OAM inside proton \cite{Filippone:2001ux,Kuhn:2008sy,Liu:2015xha}.
It is extremely challenging to achieve this in practice because it requires measuring GPD for all $x$ at fixed $\xi$.
Nevertheless, future electron-ion colliders worldwide have great potential to advance our knowledge of quark OAM and also proton tomography. It is therefore crucial to obtain remarkably various polarization choices and wide kinematic reach with the help of high luminosity accelerators and hermetic detectors.
Though global GPD fits over the whole DVCS kinematic domain, from the glue to the valence region, have remained intractable yet,
the local and global fits of CFFs are available at the amplitude level.
These tools motivate a feasibility study for the determination of CFFs by versatile observables at EicC and other facilities,
and provide one of the cornerstones of the machine construction.
In return, the design of the EicC offers unprecedented new opportunities to inspect and ameliorate these fits by feed them with pseudo-data at sea quark region.

The asymmetry measurement of DVCS with transversely polarized proton beam is particularly relevant for the experimental determination of the parton OAM.
A selected modulation of $\textrm{A}_{\textrm{UT}}$ has the best sensitivity to E to the imaginary part of CFF ${\mathcal E}$.
The DVCS simulations used for our studies are based on the physical Ansatz of constant Regge slope $B$,
and the projected events at EicC are selected in the appropriate kinematic region,
covering a domain of $1.0$ GeV$^2 < Q^2 <80.0$ GeV$^2$, 0.004 $< \xbj <$ 0.3, and $|t| >$ 0.002 GeV$^2$.
The absolute statistical uncertainty for the measured asymmetry can be as low as 0.01 for $|t| >$ 0.01 GeV$^2$.
The relative uncertainties below a few percentage, judged by the magnitude of asymmetry in GK models.
The uncertainty for the extraction of the $\textmd{Im} {\mathcal E}$ is unambiguously reduced around the sea region as long as the asymmetry pseudo-data of EicC are included into the reweighting procedure. 
Meanwhile we take advantage of high $Q^2$ region to disentangle the constraining power of $\textmd{Im} {\mathcal E}$
since GPDs are also subject to QCD evolution.
This serves as the best check so far of the relationship of DVCS single spin asymmetry and CFFs in a non-local fashion,
and reveals future experimental constraints on this observable.
Measurements of the $\textrm{A}_{\textrm{UT}}$ asymmetry must be completed to realise the full physics potential of EicC.

{At last}, let us highlight some specific perspective of future evolvement.
The local extraction of CFFs is another frequently used alternative for propagating uncertainties of data to amplitudes~\cite{Kumericki:2013br,Shiells:2021xqo}.
A recent progress toward Rosenbluth extraction framework for CFF has captured the essentials of DVCS architecture but still needs to be solidified~\cite{Kriesten:2019jep,Kriesten:2020apm,Kriesten:2020wcx}.
Moreover, more efforts are required to extrapolate zero skewness of GPDs by theoretical calculation, \emph{.e.g.} in nonlocal chiral effective theory~\cite{He:2022leb} and basis light-front quantization \cite{Xu:2021wwj,Liu:2022fvl}, to the skewness covered by the facilities.
Fortunately, lattice QCD calculations of GPDs have made rapid progress~\cite{Lin:2017snn,Lin:2020rut,Lin:2021brq,Lin:2020rxa,Alexandrou:2020zbe,Alexandrou:2021bbo,Bhattacharya:2021oyr,CSSMQCDSFUKQCD:2021lkf,Engelhardt:2017miy,Engelhardt:2020qtg}.
During the development of these tools, it will become apparent that one must measure all of the DVCS observables at several facilities to obtain continuous kinematic coverage across the large $\xbj$ down to the saturation regime
that suffice for a complete 3D images of proton in the language of GPDs.
With a complete knowledge of these functions it will be possible to quest for instance the OAM and proton spin puzzle through Ji's sum rules.
Last but not least we address that
the polarized light-ion beams such as $^3$He with a polarization of 70\% at EicC and EIC
potentially allows one to separate the quark flavors by DVCS off the neutron~\cite{Benali:2020vma,Cuic:2020iwt} together with hard exclusive meson electroproduction (DEMP)~\cite{Ellinghaus:2005uc} and the positron beam at JLab~\cite{Dutrieux:2021ehx}.


\begin{acknowledgments}

We would like to thank EicC community for the useful and inspiring discussions and PARTONS for the granted access to code.
We are grateful to Jianping Chen, Herv\'e Moutarde, Yutie Liang, Tianbo Liu, K. Kumeri\v{c}ki, Nu Xu, Zhihong Ye, Yuxiang Zhao, and Jian Zhou for their continuous support.
Special thanks to Valerio Bertone and Pawe{\l} Sznajder for a careful polish of the manuscript.
We acknowledge Xurong Chen, Qiang Fu, and Zhiwen Zhao for their contributions to this paper in the early stages of its development.
This work was supported by the National Natural Science Foundation of China (Grants Nos. 12075289 and U2032109) and the Strategic Priority Research Program of Chinese Academy of Sciences (Grant NO. XDB34030301). JZ is supported by the Qilu Youth Scholar Funding of Shandong University.

\end{acknowledgments}

\textbf{Data Availability Statement}
This manuscript has no associated data or the data will not be deposited. [Authors' comment:The pseudo-datasets generated during and/or analysed during the current study are available from the corresponding author on reasonable request.]


\bibliography{CFFatEicC.bbl}

\begin{thebibliography}{116}
\expandafter\ifx\csname natexlab\endcsname\relax\def\natexlab#1{#1}\fi
\expandafter\ifx\csname bibnamefont\endcsname\relax
  \def\bibnamefont#1{#1}\fi
\expandafter\ifx\csname bibfnamefont\endcsname\relax
  \def\bibfnamefont#1{#1}\fi
\expandafter\ifx\csname citenamefont\endcsname\relax
  \def\citenamefont#1{#1}\fi
\expandafter\ifx\csname url\endcsname\relax
  \def\url#1{\texttt{#1}}\fi
\expandafter\ifx\csname urlprefix\endcsname\relax\def\urlprefix{URL }\fi
\providecommand{\bibinfo}[2]{#2}
\providecommand{\eprint}[2][]{\url{#2}}

\bibitem[{\citenamefont{M\"uller et~al.}(1994)\citenamefont{M\"uller,
  Robaschik, Geyer, Dittes, and Ho\v{r}ej\v{s}i}}]{Mueller:1998fv}
\bibinfo{author}{\bibfnamefont{D.}~\bibnamefont{M\"uller}},
  \bibinfo{author}{\bibfnamefont{D.}~\bibnamefont{Robaschik}},
  \bibinfo{author}{\bibfnamefont{B.}~\bibnamefont{Geyer}},
  \bibinfo{author}{\bibfnamefont{F.~M.} \bibnamefont{Dittes}},
  \bibnamefont{and}
  \bibinfo{author}{\bibfnamefont{J.}~\bibnamefont{Ho\v{r}ej\v{s}i}},
  \bibinfo{journal}{Fortsch. Phys.} \textbf{\bibinfo{volume}{42}},
  \bibinfo{pages}{101} (\bibinfo{year}{1994}), \eprint{hep-ph/9812448}.

\bibitem[{\citenamefont{Ji}(1997{\natexlab{a}})}]{Ji:1996nm}
\bibinfo{author}{\bibfnamefont{X.-D.} \bibnamefont{Ji}},
  \bibinfo{journal}{Phys. Rev. D} \textbf{\bibinfo{volume}{55}},
  \bibinfo{pages}{7114} (\bibinfo{year}{1997}{\natexlab{a}}),
  \eprint{hep-ph/9609381}.

\bibitem[{\citenamefont{Ji}(1997{\natexlab{b}})}]{Ji:1996ek}
\bibinfo{author}{\bibfnamefont{X.-D.} \bibnamefont{Ji}},
  \bibinfo{journal}{Phys. Rev. Lett.} \textbf{\bibinfo{volume}{78}},
  \bibinfo{pages}{610} (\bibinfo{year}{1997}{\natexlab{b}}),
  \eprint{hep-ph/9603249}.

\bibitem[{\citenamefont{Radyushkin}(1996)}]{Radyushkin:1996nd}
\bibinfo{author}{\bibfnamefont{A.~V.} \bibnamefont{Radyushkin}},
  \bibinfo{journal}{Phys. Lett. B} \textbf{\bibinfo{volume}{380}},
  \bibinfo{pages}{417} (\bibinfo{year}{1996}), \eprint{hep-ph/9604317}.

\bibitem[{\citenamefont{Radyushkin}(1997)}]{Radyushkin:1997ki}
\bibinfo{author}{\bibfnamefont{A.~V.} \bibnamefont{Radyushkin}},
  \bibinfo{journal}{Phys. Rev. D} \textbf{\bibinfo{volume}{56}},
  \bibinfo{pages}{5524} (\bibinfo{year}{1997}), \eprint{hep-ph/9704207}.

\bibitem[{\citenamefont{Ji}(1998)}]{Ji:1998pc}
\bibinfo{author}{\bibfnamefont{X.-D.} \bibnamefont{Ji}}, \bibinfo{journal}{J.
  Phys. G} \textbf{\bibinfo{volume}{24}}, \bibinfo{pages}{1181}
  (\bibinfo{year}{1998}), \eprint{hep-ph/9807358}.

\bibitem[{\citenamefont{Diehl}(2003)}]{Diehl:2003ny}
\bibinfo{author}{\bibfnamefont{M.}~\bibnamefont{Diehl}},
  \bibinfo{journal}{Phys. Rept.} \textbf{\bibinfo{volume}{388}},
  \bibinfo{pages}{41} (\bibinfo{year}{2003}), \eprint{hep-ph/0307382}.

\bibitem[{\citenamefont{Belitsky and Radyushkin}(2005)}]{Belitsky:2005qn}
\bibinfo{author}{\bibfnamefont{A.~V.} \bibnamefont{Belitsky}} \bibnamefont{and}
  \bibinfo{author}{\bibfnamefont{A.~V.} \bibnamefont{Radyushkin}},
  \bibinfo{journal}{Phys. Rept.} \textbf{\bibinfo{volume}{418}},
  \bibinfo{pages}{1} (\bibinfo{year}{2005}), \eprint{hep-ph/0504030}.

\bibitem[{\citenamefont{Guidal et~al.}(2013)\citenamefont{Guidal, Moutarde, and
  Vanderhaeghen}}]{Guidal:2013rya}
\bibinfo{author}{\bibfnamefont{M.}~\bibnamefont{Guidal}},
  \bibinfo{author}{\bibfnamefont{H.}~\bibnamefont{Moutarde}}, \bibnamefont{and}
  \bibinfo{author}{\bibfnamefont{M.}~\bibnamefont{Vanderhaeghen}},
  \bibinfo{journal}{Rept. Prog. Phys.} \textbf{\bibinfo{volume}{76}},
  \bibinfo{pages}{066202} (\bibinfo{year}{2013}), \eprint{1303.6600}.

\bibitem[{\citenamefont{Kumericki et~al.}(2011)\citenamefont{Kumericki,
  Mueller, and Schafer}}]{Kumericki:2011rz}
\bibinfo{author}{\bibfnamefont{K.}~\bibnamefont{Kumericki}},
  \bibinfo{author}{\bibfnamefont{D.}~\bibnamefont{Mueller}}, \bibnamefont{and}
  \bibinfo{author}{\bibfnamefont{A.}~\bibnamefont{Schafer}},
  \bibinfo{journal}{JHEP} \textbf{\bibinfo{volume}{07}}, \bibinfo{pages}{073}
  (\bibinfo{year}{2011}), \eprint{1106.2808}.

\bibitem[{\citenamefont{\v{C}ui\'c et~al.}(2020)\citenamefont{\v{C}ui\'c,
  Kumeri\v{c}ki, and Sch\"afer}}]{Cuic:2020iwt}
\bibinfo{author}{\bibfnamefont{M.}~\bibnamefont{\v{C}ui\'c}},
  \bibinfo{author}{\bibfnamefont{K.}~\bibnamefont{Kumeri\v{c}ki}},
  \bibnamefont{and}
  \bibinfo{author}{\bibfnamefont{A.}~\bibnamefont{Sch\"afer}},
  \bibinfo{journal}{Phys. Rev. Lett.} \textbf{\bibinfo{volume}{125}},
  \bibinfo{pages}{232005} (\bibinfo{year}{2020}), \eprint{2007.00029}.

\bibitem[{\citenamefont{Berthou et~al.}(2018)}]{Berthou:2015oaw}
\bibinfo{author}{\bibfnamefont{B.}~\bibnamefont{Berthou}} \bibnamefont{et~al.},
  \bibinfo{journal}{Eur. Phys. J. C} \textbf{\bibinfo{volume}{78}},
  \bibinfo{pages}{478} (\bibinfo{year}{2018}), \eprint{1512.06174}.

\bibitem[{\citenamefont{Moutarde et~al.}(2019)\citenamefont{Moutarde, Sznajder,
  and Wagner}}]{Moutarde:2019tqa}
\bibinfo{author}{\bibfnamefont{H.}~\bibnamefont{Moutarde}},
  \bibinfo{author}{\bibfnamefont{P.}~\bibnamefont{Sznajder}}, \bibnamefont{and}
  \bibinfo{author}{\bibfnamefont{J.}~\bibnamefont{Wagner}},
  \bibinfo{journal}{Eur. Phys. J. C} \textbf{\bibinfo{volume}{79}},
  \bibinfo{pages}{614} (\bibinfo{year}{2019}), \eprint{1905.02089}.

\bibitem[{\citenamefont{Grigsby et~al.}(2021)\citenamefont{Grigsby, Kriesten,
  Hoskins, Liuti, Alonzi, and Burkardt}}]{Grigsby:2020auv}
\bibinfo{author}{\bibfnamefont{J.}~\bibnamefont{Grigsby}},
  \bibinfo{author}{\bibfnamefont{B.}~\bibnamefont{Kriesten}},
  \bibinfo{author}{\bibfnamefont{J.}~\bibnamefont{Hoskins}},
  \bibinfo{author}{\bibfnamefont{S.}~\bibnamefont{Liuti}},
  \bibinfo{author}{\bibfnamefont{P.}~\bibnamefont{Alonzi}}, \bibnamefont{and}
  \bibinfo{author}{\bibfnamefont{M.}~\bibnamefont{Burkardt}},
  \bibinfo{journal}{Phys. Rev. D} \textbf{\bibinfo{volume}{104}},
  \bibinfo{pages}{016001} (\bibinfo{year}{2021}), \eprint{2012.04801}.

\bibitem[{\citenamefont{Kumericki et~al.}(2008)\citenamefont{Kumericki,
  Mueller, and Passek-Kumericki}}]{Kumericki:2007sa}
\bibinfo{author}{\bibfnamefont{K.}~\bibnamefont{Kumericki}},
  \bibinfo{author}{\bibfnamefont{D.}~\bibnamefont{Mueller}}, \bibnamefont{and}
  \bibinfo{author}{\bibfnamefont{K.}~\bibnamefont{Passek-Kumericki}},
  \bibinfo{journal}{Nucl. Phys. B} \textbf{\bibinfo{volume}{794}},
  \bibinfo{pages}{244} (\bibinfo{year}{2008}), \eprint{hep-ph/0703179}.

\bibitem[{\citenamefont{Kumeri\v{c}ki and Mueller}(2010)}]{Kumericki:2009uq}
\bibinfo{author}{\bibfnamefont{K.}~\bibnamefont{Kumeri\v{c}ki}}
  \bibnamefont{and} \bibinfo{author}{\bibfnamefont{D.}~\bibnamefont{Mueller}},
  \bibinfo{journal}{Nucl. Phys. B} \textbf{\bibinfo{volume}{841}},
  \bibinfo{pages}{1} (\bibinfo{year}{2010}), \eprint{0904.0458}.

\bibitem[{\citenamefont{Moutarde et~al.}(2018)\citenamefont{Moutarde, Sznajder,
  and Wagner}}]{Moutarde:2018kwr}
\bibinfo{author}{\bibfnamefont{H.}~\bibnamefont{Moutarde}},
  \bibinfo{author}{\bibfnamefont{P.}~\bibnamefont{Sznajder}}, \bibnamefont{and}
  \bibinfo{author}{\bibfnamefont{J.}~\bibnamefont{Wagner}},
  \bibinfo{journal}{Eur. Phys. J. C} \textbf{\bibinfo{volume}{78}},
  \bibinfo{pages}{890} (\bibinfo{year}{2018}), \eprint{1807.07620}.

\bibitem[{\citenamefont{Kriesten
  et~al.}(2022{\natexlab{a}})\citenamefont{Kriesten, Velie, Yeats, Lopez, and
  Liuti}}]{Kriesten:2021sqc}
\bibinfo{author}{\bibfnamefont{B.}~\bibnamefont{Kriesten}},
  \bibinfo{author}{\bibfnamefont{P.}~\bibnamefont{Velie}},
  \bibinfo{author}{\bibfnamefont{E.}~\bibnamefont{Yeats}},
  \bibinfo{author}{\bibfnamefont{F.~Y.} \bibnamefont{Lopez}}, \bibnamefont{and}
  \bibinfo{author}{\bibfnamefont{S.}~\bibnamefont{Liuti}},
  \bibinfo{journal}{Phys. Rev. D} \textbf{\bibinfo{volume}{105}},
  \bibinfo{pages}{056022} (\bibinfo{year}{2022}{\natexlab{a}}),
  \eprint{2101.01826}.

\bibitem[{\citenamefont{Bertone et~al.}(2021)\citenamefont{Bertone, Dutrieux,
  Mezrag, Moutarde, and Sznajder}}]{Bertone:2021yyz}
\bibinfo{author}{\bibfnamefont{V.}~\bibnamefont{Bertone}},
  \bibinfo{author}{\bibfnamefont{H.}~\bibnamefont{Dutrieux}},
  \bibinfo{author}{\bibfnamefont{C.}~\bibnamefont{Mezrag}},
  \bibinfo{author}{\bibfnamefont{H.}~\bibnamefont{Moutarde}}, \bibnamefont{and}
  \bibinfo{author}{\bibfnamefont{P.}~\bibnamefont{Sznajder}},
  \bibinfo{journal}{Phys. Rev. D} \textbf{\bibinfo{volume}{103}},
  \bibinfo{pages}{114019} (\bibinfo{year}{2021}), \eprint{2104.03836}.

\bibitem[{\citenamefont{Dutrieux et~al.}(2022)\citenamefont{Dutrieux, Dutrieux,
  Grocholski, Grocholski, Moutarde, Moutarde, Sznajder, and
  Sznajder}}]{Dutrieux:2021wll}
\bibinfo{author}{\bibfnamefont{H.}~\bibnamefont{Dutrieux}},
  \bibinfo{author}{\bibfnamefont{H.}~\bibnamefont{Dutrieux}},
  \bibinfo{author}{\bibfnamefont{O.}~\bibnamefont{Grocholski}},
  \bibinfo{author}{\bibfnamefont{O.}~\bibnamefont{Grocholski}},
  \bibinfo{author}{\bibfnamefont{H.}~\bibnamefont{Moutarde}},
  \bibinfo{author}{\bibfnamefont{H.}~\bibnamefont{Moutarde}},
  \bibinfo{author}{\bibfnamefont{P.}~\bibnamefont{Sznajder}}, \bibnamefont{and}
  \bibinfo{author}{\bibfnamefont{P.}~\bibnamefont{Sznajder}},
  \bibinfo{journal}{Eur. Phys. J. C} \textbf{\bibinfo{volume}{82}},
  \bibinfo{pages}{252} (\bibinfo{year}{2022}), \bibinfo{note}{[Erratum:
  Eur.Phys.J.C 82, 389 (2022)]}, \eprint{2112.10528}.

\bibitem[{\citenamefont{Polyakov}(2003)}]{Polyakov:2002yz}
\bibinfo{author}{\bibfnamefont{M.~V.} \bibnamefont{Polyakov}},
  \bibinfo{journal}{Phys. Lett. B} \textbf{\bibinfo{volume}{555}},
  \bibinfo{pages}{57} (\bibinfo{year}{2003}), \eprint{hep-ph/0210165}.

\bibitem[{\citenamefont{Goeke et~al.}(2007)\citenamefont{Goeke, Grabis,
  Ossmann, Polyakov, Schweitzer, Silva, and Urbano}}]{Goeke:2007fp}
\bibinfo{author}{\bibfnamefont{K.}~\bibnamefont{Goeke}},
  \bibinfo{author}{\bibfnamefont{J.}~\bibnamefont{Grabis}},
  \bibinfo{author}{\bibfnamefont{J.}~\bibnamefont{Ossmann}},
  \bibinfo{author}{\bibfnamefont{M.~V.} \bibnamefont{Polyakov}},
  \bibinfo{author}{\bibfnamefont{P.}~\bibnamefont{Schweitzer}},
  \bibinfo{author}{\bibfnamefont{A.}~\bibnamefont{Silva}}, \bibnamefont{and}
  \bibinfo{author}{\bibfnamefont{D.}~\bibnamefont{Urbano}},
  \bibinfo{journal}{Phys. Rev. D} \textbf{\bibinfo{volume}{75}},
  \bibinfo{pages}{094021} (\bibinfo{year}{2007}), \eprint{hep-ph/0702030}.

\bibitem[{\citenamefont{Lorc\'e et~al.}(2019)\citenamefont{Lorc\'e, Moutarde,
  and Trawi\'nski}}]{Lorce:2018egm}
\bibinfo{author}{\bibfnamefont{C.}~\bibnamefont{Lorc\'e}},
  \bibinfo{author}{\bibfnamefont{H.}~\bibnamefont{Moutarde}}, \bibnamefont{and}
  \bibinfo{author}{\bibfnamefont{A.~P.} \bibnamefont{Trawi\'nski}},
  \bibinfo{journal}{Eur. Phys. J. C} \textbf{\bibinfo{volume}{79}},
  \bibinfo{pages}{89} (\bibinfo{year}{2019}), \eprint{1810.09837}.

\bibitem[{\citenamefont{Polyakov and Schweitzer}(2018)}]{Polyakov:2018zvc}
\bibinfo{author}{\bibfnamefont{M.~V.} \bibnamefont{Polyakov}} \bibnamefont{and}
  \bibinfo{author}{\bibfnamefont{P.}~\bibnamefont{Schweitzer}},
  \bibinfo{journal}{Int. J. Mod. Phys. A} \textbf{\bibinfo{volume}{33}},
  \bibinfo{pages}{1830025} (\bibinfo{year}{2018}), \eprint{1805.06596}.

\bibitem[{\citenamefont{Shanahan and Detmold}(2019)}]{Shanahan:2018nnv}
\bibinfo{author}{\bibfnamefont{P.~E.} \bibnamefont{Shanahan}} \bibnamefont{and}
  \bibinfo{author}{\bibfnamefont{W.}~\bibnamefont{Detmold}},
  \bibinfo{journal}{Phys. Rev. Lett.} \textbf{\bibinfo{volume}{122}},
  \bibinfo{pages}{072003} (\bibinfo{year}{2019}), \eprint{1810.07589}.

\bibitem[{\citenamefont{Burkert et~al.}(2018)\citenamefont{Burkert,
  Elouadrhiri, and Girod}}]{Burkert:2018bqq}
\bibinfo{author}{\bibfnamefont{V.~D.} \bibnamefont{Burkert}},
  \bibinfo{author}{\bibfnamefont{L.}~\bibnamefont{Elouadrhiri}},
  \bibnamefont{and} \bibinfo{author}{\bibfnamefont{F.~X.} \bibnamefont{Girod}},
  \bibinfo{journal}{Nature} \textbf{\bibinfo{volume}{557}},
  \bibinfo{pages}{396} (\bibinfo{year}{2018}).

\bibitem[{\citenamefont{Kumeri\v{c}ki}(2019)}]{Kumericki:2019ddg}
\bibinfo{author}{\bibfnamefont{K.}~\bibnamefont{Kumeri\v{c}ki}},
  \bibinfo{journal}{Nature} \textbf{\bibinfo{volume}{570}}, \bibinfo{pages}{E1}
  (\bibinfo{year}{2019}).

\bibitem[{\citenamefont{Chekanov et~al.}(2003)}]{Chekanov:2003ya}
\bibinfo{author}{\bibfnamefont{S.}~\bibnamefont{Chekanov}} \bibnamefont{et~al.}
  (\bibinfo{collaboration}{ZEUS}), \bibinfo{journal}{Phys. Lett. B}
  \textbf{\bibinfo{volume}{573}}, \bibinfo{pages}{46} (\bibinfo{year}{2003}),
  \eprint{hep-ex/0305028}.

\bibitem[{\citenamefont{Chekanov et~al.}(2009)}]{Chekanov:2008vy}
\bibinfo{author}{\bibfnamefont{S.}~\bibnamefont{Chekanov}} \bibnamefont{et~al.}
  (\bibinfo{collaboration}{ZEUS}), \bibinfo{journal}{JHEP}
  \textbf{\bibinfo{volume}{05}}, \bibinfo{pages}{108} (\bibinfo{year}{2009}),
  \eprint{0812.2517}.

\bibitem[{\citenamefont{Adloff et~al.}(2001)}]{Adloff:2001cn}
\bibinfo{author}{\bibfnamefont{C.}~\bibnamefont{Adloff}} \bibnamefont{et~al.}
  (\bibinfo{collaboration}{H1}), \bibinfo{journal}{Phys. Lett. B}
  \textbf{\bibinfo{volume}{517}}, \bibinfo{pages}{47} (\bibinfo{year}{2001}),
  \eprint{hep-ex/0107005}.

\bibitem[{\citenamefont{Aktas et~al.}(2005)}]{Aktas:2005ty}
\bibinfo{author}{\bibfnamefont{A.}~\bibnamefont{Aktas}} \bibnamefont{et~al.}
  (\bibinfo{collaboration}{H1}), \bibinfo{journal}{Eur. Phys. J. C}
  \textbf{\bibinfo{volume}{44}}, \bibinfo{pages}{1} (\bibinfo{year}{2005}),
  \eprint{hep-ex/0505061}.

\bibitem[{\citenamefont{Aaron et~al.}(2009{\natexlab{a}})}]{H1:2009wnw}
\bibinfo{author}{\bibfnamefont{F.~D.} \bibnamefont{Aaron}} \bibnamefont{et~al.}
  (\bibinfo{collaboration}{H1}), \bibinfo{journal}{Phys. Lett. B}
  \textbf{\bibinfo{volume}{681}}, \bibinfo{pages}{391}
  (\bibinfo{year}{2009}{\natexlab{a}}), \eprint{0907.5289}.

\bibitem[{\citenamefont{Aaron et~al.}(2009{\natexlab{b}})}]{Aaron:2009ac}
\bibinfo{author}{\bibfnamefont{F.~D.} \bibnamefont{Aaron}} \bibnamefont{et~al.}
  (\bibinfo{collaboration}{H1}), \bibinfo{journal}{Phys. Lett. B}
  \textbf{\bibinfo{volume}{681}}, \bibinfo{pages}{391}
  (\bibinfo{year}{2009}{\natexlab{b}}), \eprint{0907.5289}.

\bibitem[{\citenamefont{Airapetian et~al.}(2012)}]{Airapetian:2012pg}
\bibinfo{author}{\bibfnamefont{A.}~\bibnamefont{Airapetian}}
  \bibnamefont{et~al.} (\bibinfo{collaboration}{HERMES}),
  \bibinfo{journal}{JHEP} \textbf{\bibinfo{volume}{10}}, \bibinfo{pages}{042}
  (\bibinfo{year}{2012}), \eprint{1206.5683}.

\bibitem[{\citenamefont{Akhunzyanov et~al.}(2019)}]{COMPASS:2018pup}
\bibinfo{author}{\bibfnamefont{R.}~\bibnamefont{Akhunzyanov}}
  \bibnamefont{et~al.} (\bibinfo{collaboration}{COMPASS}),
  \bibinfo{journal}{Phys. Lett. B} \textbf{\bibinfo{volume}{793}},
  \bibinfo{pages}{188} (\bibinfo{year}{2019}), \bibinfo{note}{[Erratum:
  Phys.Lett.B 800, 135129 (2020)]}, \eprint{1802.02739}.

\bibitem[{\citenamefont{Defurne et~al.}(2017)}]{Defurne:2017paw}
\bibinfo{author}{\bibfnamefont{M.}~\bibnamefont{Defurne}} \bibnamefont{et~al.},
  \bibinfo{journal}{Nature Commun.} \textbf{\bibinfo{volume}{8}},
  \bibinfo{pages}{1408} (\bibinfo{year}{2017}), \eprint{1703.09442}.

\bibitem[{\citenamefont{Defurne et~al.}(2015)}]{Defurne:2015kxq}
\bibinfo{author}{\bibfnamefont{M.}~\bibnamefont{Defurne}} \bibnamefont{et~al.}
  (\bibinfo{collaboration}{Jefferson Lab Hall A}), \bibinfo{journal}{Phys. Rev.
  C} \textbf{\bibinfo{volume}{92}}, \bibinfo{pages}{055202}
  (\bibinfo{year}{2015}), \eprint{1504.05453}.

\bibitem[{\citenamefont{d'Hose et~al.}(2016)\citenamefont{d'Hose, Niccolai, and
  Rostomyan}}]{dHose:2016mda}
\bibinfo{author}{\bibfnamefont{N.}~\bibnamefont{d'Hose}},
  \bibinfo{author}{\bibfnamefont{S.}~\bibnamefont{Niccolai}}, \bibnamefont{and}
  \bibinfo{author}{\bibfnamefont{A.}~\bibnamefont{Rostomyan}},
  \bibinfo{journal}{Eur. Phys. J. A} \textbf{\bibinfo{volume}{52}},
  \bibinfo{pages}{151} (\bibinfo{year}{2016}).

\bibitem[{\citenamefont{Hyde et~al.}(2011)\citenamefont{Hyde, Guidal, and
  Radyushkin}}]{Hyde:2011ke}
\bibinfo{author}{\bibfnamefont{C.~E.} \bibnamefont{Hyde}},
  \bibinfo{author}{\bibfnamefont{M.}~\bibnamefont{Guidal}}, \bibnamefont{and}
  \bibinfo{author}{\bibfnamefont{A.~V.} \bibnamefont{Radyushkin}},
  \bibinfo{journal}{J. Phys. Conf. Ser.} \textbf{\bibinfo{volume}{299}},
  \bibinfo{pages}{012006} (\bibinfo{year}{2011}), \eprint{1101.2482}.

\bibitem[{\citenamefont{Burkert et~al.}(2021)}]{Burkert:2021rxz}
\bibinfo{author}{\bibfnamefont{V.}~\bibnamefont{Burkert}} \bibnamefont{et~al.}
  (\bibinfo{collaboration}{CLAS}), \bibinfo{journal}{Eur. Phys. J. A}
  \textbf{\bibinfo{volume}{57}}, \bibinfo{pages}{186} (\bibinfo{year}{2021}),
  \eprint{2103.12651}.

\bibitem[{\citenamefont{Fucini et~al.}(2021)\citenamefont{Fucini, Hattawy,
  Rinaldi, and Scopetta}}]{Fucini:2021psq}
\bibinfo{author}{\bibfnamefont{S.}~\bibnamefont{Fucini}},
  \bibinfo{author}{\bibfnamefont{M.}~\bibnamefont{Hattawy}},
  \bibinfo{author}{\bibfnamefont{M.}~\bibnamefont{Rinaldi}}, \bibnamefont{and}
  \bibinfo{author}{\bibfnamefont{S.}~\bibnamefont{Scopetta}},
  \bibinfo{journal}{Eur. Phys. J. A} \textbf{\bibinfo{volume}{57}},
  \bibinfo{pages}{273} (\bibinfo{year}{2021}), \eprint{2105.00435}.

\bibitem[{\citenamefont{Afanasev et~al.}(2021)}]{Afanasev:2021twk}
\bibinfo{author}{\bibfnamefont{A.}~\bibnamefont{Afanasev}}
  \bibnamefont{et~al.}, \bibinfo{journal}{Eur. Phys. J. A}
  \textbf{\bibinfo{volume}{57}}, \bibinfo{pages}{300} (\bibinfo{year}{2021}),
  \eprint{2105.06540}.

\bibitem[{\citenamefont{Deshpande et~al.}(2005)\citenamefont{Deshpande, Milner,
  Venugopalan, and Vogelsang}}]{Deshpande:2005wd}
\bibinfo{author}{\bibfnamefont{A.}~\bibnamefont{Deshpande}},
  \bibinfo{author}{\bibfnamefont{R.}~\bibnamefont{Milner}},
  \bibinfo{author}{\bibfnamefont{R.}~\bibnamefont{Venugopalan}},
  \bibnamefont{and}
  \bibinfo{author}{\bibfnamefont{W.}~\bibnamefont{Vogelsang}},
  \bibinfo{journal}{Ann. Rev. Nucl. Part. Sci.} \textbf{\bibinfo{volume}{55}},
  \bibinfo{pages}{165} (\bibinfo{year}{2005}), \eprint{hep-ph/0506148}.

\bibitem[{\citenamefont{Accardi et~al.}(2016)}]{Accardi:2012qut}
\bibinfo{author}{\bibfnamefont{A.}~\bibnamefont{Accardi}} \bibnamefont{et~al.},
  \bibinfo{journal}{Eur. Phys. J. A} \textbf{\bibinfo{volume}{52}},
  \bibinfo{pages}{268} (\bibinfo{year}{2016}), \eprint{1212.1701}.

\bibitem[{\citenamefont{Abdul~Khalek et~al.}(2022)}]{AbdulKhalek:2021gbh}
\bibinfo{author}{\bibfnamefont{R.}~\bibnamefont{Abdul~Khalek}}
  \bibnamefont{et~al.}, \bibinfo{journal}{Nucl. Phys. A}
  \textbf{\bibinfo{volume}{1026}}, \bibinfo{pages}{122447}
  (\bibinfo{year}{2022}), \eprint{2103.05419}.

\bibitem[{\citenamefont{Cao et~al.}(2020{\natexlab{a}})\citenamefont{Cao,
  Chang, Chang et~al.}}]{CAO:2020EicC}
\bibinfo{author}{\bibfnamefont{X.}~\bibnamefont{Cao}},
  \bibinfo{author}{\bibfnamefont{L.}~\bibnamefont{Chang}},
  \bibinfo{author}{\bibfnamefont{N.}~\bibnamefont{Chang}},
  \bibnamefont{et~al.}, \bibinfo{journal}{Nuclear Techniques}
  \textbf{\bibinfo{volume}{43}}, \bibinfo{pages}{020001}
  (\bibinfo{year}{2020}{\natexlab{a}}).

\bibitem[{\citenamefont{Cao et~al.}(2020{\natexlab{b}})\citenamefont{Cao, Chen,
  Gong et~al.}}]{CAO:2020Sci}
\bibinfo{author}{\bibfnamefont{X.}~\bibnamefont{Cao}},
  \bibinfo{author}{\bibfnamefont{X.}~\bibnamefont{Chen}},
  \bibinfo{author}{\bibfnamefont{C.}~\bibnamefont{Gong}}, \bibnamefont{et~al.},
  \bibinfo{journal}{SCIENTIA SINICA Physica, Mechanica, Astronomica}
  \textbf{\bibinfo{volume}{50}}, \bibinfo{pages}{112005}
  (\bibinfo{year}{2020}{\natexlab{b}}).

\bibitem[{\citenamefont{Anderle et~al.}(2021)}]{Anderle:2021wcy}
\bibinfo{author}{\bibfnamefont{D.~P.} \bibnamefont{Anderle}}
  \bibnamefont{et~al.}, \bibinfo{journal}{Front. Phys. (Beijing)}
  \textbf{\bibinfo{volume}{16}}, \bibinfo{pages}{64701} (\bibinfo{year}{2021}),
  \eprint{2102.09222}.

\bibitem[{\citenamefont{Vanderhaeghen et~al.}(1999)\citenamefont{Vanderhaeghen,
  Guichon, and Guidal}}]{Vanderhaeghen:1999xj}
\bibinfo{author}{\bibfnamefont{M.}~\bibnamefont{Vanderhaeghen}},
  \bibinfo{author}{\bibfnamefont{P.~A.~M.} \bibnamefont{Guichon}},
  \bibnamefont{and} \bibinfo{author}{\bibfnamefont{M.}~\bibnamefont{Guidal}},
  \bibinfo{journal}{Phys. Rev. D} \textbf{\bibinfo{volume}{60}},
  \bibinfo{pages}{094017} (\bibinfo{year}{1999}), \eprint{hep-ph/9905372}.

\bibitem[{\citenamefont{Dupre et~al.}(2017)\citenamefont{Dupre, Guidal, and
  Vanderhaeghen}}]{Dupre:2016mai}
\bibinfo{author}{\bibfnamefont{R.}~\bibnamefont{Dupre}},
  \bibinfo{author}{\bibfnamefont{M.}~\bibnamefont{Guidal}}, \bibnamefont{and}
  \bibinfo{author}{\bibfnamefont{M.}~\bibnamefont{Vanderhaeghen}},
  \bibinfo{journal}{Phys. Rev. D} \textbf{\bibinfo{volume}{95}},
  \bibinfo{pages}{011501} (\bibinfo{year}{2017}), \eprint{1606.07821}.

\bibitem[{\citenamefont{Goloskokov and Kroll}(2005)}]{Goloskokov:2005sd}
\bibinfo{author}{\bibfnamefont{S.~V.} \bibnamefont{Goloskokov}}
  \bibnamefont{and} \bibinfo{author}{\bibfnamefont{P.}~\bibnamefont{Kroll}},
  \bibinfo{journal}{Eur. Phys. J. C} \textbf{\bibinfo{volume}{42}},
  \bibinfo{pages}{281} (\bibinfo{year}{2005}), \eprint{hep-ph/0501242}.

\bibitem[{\citenamefont{Goloskokov and Kroll}(2007)}]{Goloskokov:2006hr}
\bibinfo{author}{\bibfnamefont{S.~V.} \bibnamefont{Goloskokov}}
  \bibnamefont{and} \bibinfo{author}{\bibfnamefont{P.}~\bibnamefont{Kroll}},
  \bibinfo{journal}{Eur. Phys. J. C} \textbf{\bibinfo{volume}{50}},
  \bibinfo{pages}{829} (\bibinfo{year}{2007}), \eprint{hep-ph/0611290}.

\bibitem[{\citenamefont{Goloskokov and Kroll}(2008)}]{Goloskokov:2007nt}
\bibinfo{author}{\bibfnamefont{S.~V.} \bibnamefont{Goloskokov}}
  \bibnamefont{and} \bibinfo{author}{\bibfnamefont{P.}~\bibnamefont{Kroll}},
  \bibinfo{journal}{Eur. Phys. J. C} \textbf{\bibinfo{volume}{53}},
  \bibinfo{pages}{367} (\bibinfo{year}{2008}), \eprint{0708.3569}.

\bibitem[{\citenamefont{Goloskokov and Kroll}(2009)}]{Goloskokov:2008ib}
\bibinfo{author}{\bibfnamefont{S.~V.} \bibnamefont{Goloskokov}}
  \bibnamefont{and} \bibinfo{author}{\bibfnamefont{P.}~\bibnamefont{Kroll}},
  \bibinfo{journal}{Eur. Phys. J. C} \textbf{\bibinfo{volume}{59}},
  \bibinfo{pages}{809} (\bibinfo{year}{2009}), \eprint{0809.4126}.

\bibitem[{\citenamefont{Goloskokov and Kroll}(2010)}]{Goloskokov:2009ia}
\bibinfo{author}{\bibfnamefont{S.~V.} \bibnamefont{Goloskokov}}
  \bibnamefont{and} \bibinfo{author}{\bibfnamefont{P.}~\bibnamefont{Kroll}},
  \bibinfo{journal}{Eur. Phys. J. C} \textbf{\bibinfo{volume}{65}},
  \bibinfo{pages}{137} (\bibinfo{year}{2010}), \eprint{0906.0460}.

\bibitem[{\citenamefont{Goloskokov and Kroll}(2011)}]{Goloskokov:2011rd}
\bibinfo{author}{\bibfnamefont{S.~V.} \bibnamefont{Goloskokov}}
  \bibnamefont{and} \bibinfo{author}{\bibfnamefont{P.}~\bibnamefont{Kroll}},
  \bibinfo{journal}{Eur. Phys. J. A} \textbf{\bibinfo{volume}{47}},
  \bibinfo{pages}{112} (\bibinfo{year}{2011}), \eprint{1106.4897}.

\bibitem[{\citenamefont{Goloskokov and
  Kroll}(2014{\natexlab{a}})}]{Goloskokov:2013mba}
\bibinfo{author}{\bibfnamefont{S.~V.} \bibnamefont{Goloskokov}}
  \bibnamefont{and} \bibinfo{author}{\bibfnamefont{P.}~\bibnamefont{Kroll}},
  \bibinfo{journal}{Eur. Phys. J. C} \textbf{\bibinfo{volume}{74}},
  \bibinfo{pages}{2725} (\bibinfo{year}{2014}{\natexlab{a}}),
  \eprint{1310.1472}.

\bibitem[{\citenamefont{Goloskokov and
  Kroll}(2014{\natexlab{b}})}]{Goloskokov:2014ika}
\bibinfo{author}{\bibfnamefont{S.~V.} \bibnamefont{Goloskokov}}
  \bibnamefont{and} \bibinfo{author}{\bibfnamefont{P.}~\bibnamefont{Kroll}},
  \bibinfo{journal}{Eur. Phys. J. A} \textbf{\bibinfo{volume}{50}},
  \bibinfo{pages}{146} (\bibinfo{year}{2014}{\natexlab{b}}),
  \eprint{1407.1141}.

\bibitem[{\citenamefont{Ye}(2006)}]{Ye:2006pe}
\bibinfo{author}{\bibfnamefont{Z.}~\bibnamefont{Ye}}, Ph.D. thesis,
  \bibinfo{school}{Hamburg U.} (\bibinfo{year}{2006}).

\bibitem[{\citenamefont{Murray}(2007)}]{Murray:2007zzb}
\bibinfo{author}{\bibfnamefont{M.~J.} \bibnamefont{Murray}}, Ph.D. thesis,
  \bibinfo{school}{Glasgow U.} (\bibinfo{year}{2007}).

\bibitem[{\citenamefont{Airapetian et~al.}(2008)}]{Airapetian:2008aa}
\bibinfo{author}{\bibfnamefont{A.}~\bibnamefont{Airapetian}}
  \bibnamefont{et~al.} (\bibinfo{collaboration}{HERMES}),
  \bibinfo{journal}{JHEP} \textbf{\bibinfo{volume}{06}}, \bibinfo{pages}{066}
  (\bibinfo{year}{2008}), \eprint{0802.2499}.

\bibitem[{\citenamefont{Airapetian et~al.}(2011)}]{Airapetian:2011uq}
\bibinfo{author}{\bibfnamefont{A.}~\bibnamefont{Airapetian}}
  \bibnamefont{et~al.} (\bibinfo{collaboration}{HERMES}),
  \bibinfo{journal}{Phys. Lett. B} \textbf{\bibinfo{volume}{704}},
  \bibinfo{pages}{15} (\bibinfo{year}{2011}), \eprint{1106.2990}.

\bibitem[{\citenamefont{Diehl and Kroll}(2013)}]{Diehl:2013xca}
\bibinfo{author}{\bibfnamefont{M.}~\bibnamefont{Diehl}} \bibnamefont{and}
  \bibinfo{author}{\bibfnamefont{P.}~\bibnamefont{Kroll}},
  \bibinfo{journal}{Eur. Phys. J. C} \textbf{\bibinfo{volume}{73}},
  \bibinfo{pages}{2397} (\bibinfo{year}{2013}), \eprint{1302.4604}.

\bibitem[{\citenamefont{Kroll}(2020)}]{Kroll:2020jat}
\bibinfo{author}{\bibfnamefont{P.}~\bibnamefont{Kroll}}, \bibinfo{journal}{Mod.
  Phys. Lett. A} \textbf{\bibinfo{volume}{35}}, \bibinfo{pages}{2050093}
  (\bibinfo{year}{2020}), \eprint{2001.01919}.

\bibitem[{\citenamefont{Almaeen et~al.}(2022)\citenamefont{Almaeen, Grigsby,
  Hoskins, Kriesten, Li, Lin, and Liuti}}]{Almaeen:2022imx}
\bibinfo{author}{\bibfnamefont{M.}~\bibnamefont{Almaeen}},
  \bibinfo{author}{\bibfnamefont{J.}~\bibnamefont{Grigsby}},
  \bibinfo{author}{\bibfnamefont{J.}~\bibnamefont{Hoskins}},
  \bibinfo{author}{\bibfnamefont{B.}~\bibnamefont{Kriesten}},
  \bibinfo{author}{\bibfnamefont{Y.}~\bibnamefont{Li}},
  \bibinfo{author}{\bibfnamefont{H.-W.} \bibnamefont{Lin}}, \bibnamefont{and}
  \bibinfo{author}{\bibfnamefont{S.}~\bibnamefont{Liuti}}
  (\bibinfo{year}{2022}), \eprint{2207.10766}.

\bibitem[{\citenamefont{Aschenauer et~al.}(2022)\citenamefont{Aschenauer,
  Batozskaya, Fazio, Gates, Moutarde, Sokhan, Spiesberger, Sznajder, and
  Tezgin}}]{Aschenauer:2022aeb}
\bibinfo{author}{\bibfnamefont{E.~C.} \bibnamefont{Aschenauer}},
  \bibinfo{author}{\bibfnamefont{V.}~\bibnamefont{Batozskaya}},
  \bibinfo{author}{\bibfnamefont{S.}~\bibnamefont{Fazio}},
  \bibinfo{author}{\bibfnamefont{K.}~\bibnamefont{Gates}},
  \bibinfo{author}{\bibfnamefont{H.}~\bibnamefont{Moutarde}},
  \bibinfo{author}{\bibfnamefont{D.}~\bibnamefont{Sokhan}},
  \bibinfo{author}{\bibfnamefont{H.}~\bibnamefont{Spiesberger}},
  \bibinfo{author}{\bibfnamefont{P.}~\bibnamefont{Sznajder}}, \bibnamefont{and}
  \bibinfo{author}{\bibfnamefont{K.}~\bibnamefont{Tezgin}},
  \bibinfo{journal}{Eur. Phys. J. C} \textbf{\bibinfo{volume}{82}},
  \bibinfo{pages}{819} (\bibinfo{year}{2022}), \eprint{2205.01762}.

\bibitem[{\citenamefont{Burkardt}(2000)}]{Burkardt:2000za}
\bibinfo{author}{\bibfnamefont{M.}~\bibnamefont{Burkardt}},
  \bibinfo{journal}{Phys. Rev. D} \textbf{\bibinfo{volume}{62}},
  \bibinfo{pages}{071503} (\bibinfo{year}{2000}), \bibinfo{note}{[Erratum:
  Phys.Rev.D 66, 119903 (2002)]}, \eprint{hep-ph/0005108}.

\bibitem[{\citenamefont{Burkardt}(2003)}]{Burkardt:2002hr}
\bibinfo{author}{\bibfnamefont{M.}~\bibnamefont{Burkardt}},
  \bibinfo{journal}{Int. J. Mod. Phys. A} \textbf{\bibinfo{volume}{18}},
  \bibinfo{pages}{173} (\bibinfo{year}{2003}), \eprint{hep-ph/0207047}.

\bibitem[{\citenamefont{Kumericki et~al.}(2016)\citenamefont{Kumericki, Liuti,
  and Moutarde}}]{Kumericki:2016ehc}
\bibinfo{author}{\bibfnamefont{K.}~\bibnamefont{Kumericki}},
  \bibinfo{author}{\bibfnamefont{S.}~\bibnamefont{Liuti}}, \bibnamefont{and}
  \bibinfo{author}{\bibfnamefont{H.}~\bibnamefont{Moutarde}},
  \bibinfo{journal}{Eur. Phys. J. A} \textbf{\bibinfo{volume}{52}},
  \bibinfo{pages}{157} (\bibinfo{year}{2016}), \eprint{1602.02763}.

\bibitem[{\citenamefont{Diehl and Sapeta}(2005)}]{Diehl:2005pc}
\bibinfo{author}{\bibfnamefont{M.}~\bibnamefont{Diehl}} \bibnamefont{and}
  \bibinfo{author}{\bibfnamefont{S.}~\bibnamefont{Sapeta}},
  \bibinfo{journal}{Eur. Phys. J. C} \textbf{\bibinfo{volume}{41}},
  \bibinfo{pages}{515} (\bibinfo{year}{2005}), \eprint{hep-ph/0503023}.

\bibitem[{\citenamefont{Aschenauer et~al.}(2012)\citenamefont{Aschenauer,
  Sassot, and Stratmann}}]{Aschenauer:2012ve}
\bibinfo{author}{\bibfnamefont{E.~C.} \bibnamefont{Aschenauer}},
  \bibinfo{author}{\bibfnamefont{R.}~\bibnamefont{Sassot}}, \bibnamefont{and}
  \bibinfo{author}{\bibfnamefont{M.}~\bibnamefont{Stratmann}},
  \bibinfo{journal}{Phys. Rev. D} \textbf{\bibinfo{volume}{86}},
  \bibinfo{pages}{054020} (\bibinfo{year}{2012}), \eprint{1206.6014}.

\bibitem[{\citenamefont{Kroll et~al.}(2013)\citenamefont{Kroll, Moutarde, and
  Sabatie}}]{Kroll:2012sm}
\bibinfo{author}{\bibfnamefont{P.}~\bibnamefont{Kroll}},
  \bibinfo{author}{\bibfnamefont{H.}~\bibnamefont{Moutarde}}, \bibnamefont{and}
  \bibinfo{author}{\bibfnamefont{F.}~\bibnamefont{Sabatie}},
  \bibinfo{journal}{Eur. Phys. J. C} \textbf{\bibinfo{volume}{73}},
  \bibinfo{pages}{2278} (\bibinfo{year}{2013}), \eprint{1210.6975}.

\bibitem[{\citenamefont{Perez et~al.}(2004)\citenamefont{Perez, Schoeffel, and
  Favart}}]{Perez:2004ig}
\bibinfo{author}{\bibfnamefont{E.}~\bibnamefont{Perez}},
  \bibinfo{author}{\bibfnamefont{L.}~\bibnamefont{Schoeffel}},
  \bibnamefont{and} \bibinfo{author}{\bibfnamefont{L.}~\bibnamefont{Favart}}
  (\bibinfo{year}{2004}), \eprint{hep-ph/0411389}.

\bibitem[{\citenamefont{Aschenauer et~al.}(2013)\citenamefont{Aschenauer,
  Fazio, Kumericki, and Mueller}}]{Aschenauer:2013hhw}
\bibinfo{author}{\bibfnamefont{E.-C.} \bibnamefont{Aschenauer}},
  \bibinfo{author}{\bibfnamefont{S.}~\bibnamefont{Fazio}},
  \bibinfo{author}{\bibfnamefont{K.}~\bibnamefont{Kumericki}},
  \bibnamefont{and} \bibinfo{author}{\bibfnamefont{D.}~\bibnamefont{Mueller}},
  \bibinfo{journal}{JHEP} \textbf{\bibinfo{volume}{09}}, \bibinfo{pages}{093}
  (\bibinfo{year}{2013}), \eprint{1304.0077}.

\bibitem[{\citenamefont{Aschenauer et~al.}(2019)\citenamefont{Aschenauer,
  Fazio, Lee, Mantysaari, Page, Schenke, Ullrich, Venugopalan, and
  Zurita}}]{Aschenauer:2017jsk}
\bibinfo{author}{\bibfnamefont{E.~C.} \bibnamefont{Aschenauer}},
  \bibinfo{author}{\bibfnamefont{S.}~\bibnamefont{Fazio}},
  \bibinfo{author}{\bibfnamefont{J.~H.} \bibnamefont{Lee}},
  \bibinfo{author}{\bibfnamefont{H.}~\bibnamefont{Mantysaari}},
  \bibinfo{author}{\bibfnamefont{B.~S.} \bibnamefont{Page}},
  \bibinfo{author}{\bibfnamefont{B.}~\bibnamefont{Schenke}},
  \bibinfo{author}{\bibfnamefont{T.}~\bibnamefont{Ullrich}},
  \bibinfo{author}{\bibfnamefont{R.}~\bibnamefont{Venugopalan}},
  \bibnamefont{and} \bibinfo{author}{\bibfnamefont{P.}~\bibnamefont{Zurita}},
  \bibinfo{journal}{Rept. Prog. Phys.} \textbf{\bibinfo{volume}{82}},
  \bibinfo{pages}{024301} (\bibinfo{year}{2019}), \eprint{1708.01527}.

\bibitem[{\citenamefont{Belitsky et~al.}(2000)\citenamefont{Belitsky, Freund,
  and Mueller}}]{Belitsky:1999hf}
\bibinfo{author}{\bibfnamefont{A.~V.} \bibnamefont{Belitsky}},
  \bibinfo{author}{\bibfnamefont{A.}~\bibnamefont{Freund}}, \bibnamefont{and}
  \bibinfo{author}{\bibfnamefont{D.}~\bibnamefont{Mueller}},
  \bibinfo{journal}{Nucl. Phys. B} \textbf{\bibinfo{volume}{574}},
  \bibinfo{pages}{347} (\bibinfo{year}{2000}), \eprint{hep-ph/9912379}.

\bibitem[{\citenamefont{Belitsky et~al.}(2002)\citenamefont{Belitsky, Mueller,
  and Kirchner}}]{Belitsky:2001ns}
\bibinfo{author}{\bibfnamefont{A.~V.} \bibnamefont{Belitsky}},
  \bibinfo{author}{\bibfnamefont{D.}~\bibnamefont{Mueller}}, \bibnamefont{and}
  \bibinfo{author}{\bibfnamefont{A.}~\bibnamefont{Kirchner}},
  \bibinfo{journal}{Nucl. Phys. B} \textbf{\bibinfo{volume}{629}},
  \bibinfo{pages}{323} (\bibinfo{year}{2002}), \eprint{hep-ph/0112108}.

\bibitem[{\citenamefont{Ball et~al.}(2011)\citenamefont{Ball, Bertone, Cerutti,
  Del~Debbio, Forte, Guffanti, Latorre, Rojo, and Ubiali}}]{Ball:2010gb}
\bibinfo{author}{\bibfnamefont{R.~D.} \bibnamefont{Ball}},
  \bibinfo{author}{\bibfnamefont{V.}~\bibnamefont{Bertone}},
  \bibinfo{author}{\bibfnamefont{F.}~\bibnamefont{Cerutti}},
  \bibinfo{author}{\bibfnamefont{L.}~\bibnamefont{Del~Debbio}},
  \bibinfo{author}{\bibfnamefont{S.}~\bibnamefont{Forte}},
  \bibinfo{author}{\bibfnamefont{A.}~\bibnamefont{Guffanti}},
  \bibinfo{author}{\bibfnamefont{J.~I.} \bibnamefont{Latorre}},
  \bibinfo{author}{\bibfnamefont{J.}~\bibnamefont{Rojo}}, \bibnamefont{and}
  \bibinfo{author}{\bibfnamefont{M.}~\bibnamefont{Ubiali}}
  (\bibinfo{collaboration}{NNPDF}), \bibinfo{journal}{Nucl. Phys. B}
  \textbf{\bibinfo{volume}{849}}, \bibinfo{pages}{112} (\bibinfo{year}{2011}),
  \bibinfo{note}{[Erratum: Nucl.Phys.B 854, 926--927 (2012), Erratum:
  Nucl.Phys.B 855, 927--928 (2012)]}, \eprint{1012.0836}.

\bibitem[{\citenamefont{Ball et~al.}(2012)\citenamefont{Ball, Bertone, Cerutti,
  Del~Debbio, Forte, Guffanti, Hartland, Latorre, Rojo, and
  Ubiali}}]{Ball:2011gg}
\bibinfo{author}{\bibfnamefont{R.~D.} \bibnamefont{Ball}},
  \bibinfo{author}{\bibfnamefont{V.}~\bibnamefont{Bertone}},
  \bibinfo{author}{\bibfnamefont{F.}~\bibnamefont{Cerutti}},
  \bibinfo{author}{\bibfnamefont{L.}~\bibnamefont{Del~Debbio}},
  \bibinfo{author}{\bibfnamefont{S.}~\bibnamefont{Forte}},
  \bibinfo{author}{\bibfnamefont{A.}~\bibnamefont{Guffanti}},
  \bibinfo{author}{\bibfnamefont{N.~P.} \bibnamefont{Hartland}},
  \bibinfo{author}{\bibfnamefont{J.~I.} \bibnamefont{Latorre}},
  \bibinfo{author}{\bibfnamefont{J.}~\bibnamefont{Rojo}}, \bibnamefont{and}
  \bibinfo{author}{\bibfnamefont{M.}~\bibnamefont{Ubiali}},
  \bibinfo{journal}{Nucl. Phys. B} \textbf{\bibinfo{volume}{855}},
  \bibinfo{pages}{608} (\bibinfo{year}{2012}), \eprint{1108.1758}.

\bibitem[{\citenamefont{Dutrieux
  et~al.}(2021{\natexlab{a}})\citenamefont{Dutrieux, Lorc\'e, Moutarde,
  Sznajder, Trawi\'nski, and Wagner}}]{Dutrieux:2021nlz}
\bibinfo{author}{\bibfnamefont{H.}~\bibnamefont{Dutrieux}},
  \bibinfo{author}{\bibfnamefont{C.}~\bibnamefont{Lorc\'e}},
  \bibinfo{author}{\bibfnamefont{H.}~\bibnamefont{Moutarde}},
  \bibinfo{author}{\bibfnamefont{P.}~\bibnamefont{Sznajder}},
  \bibinfo{author}{\bibfnamefont{A.}~\bibnamefont{Trawi\'nski}},
  \bibnamefont{and} \bibinfo{author}{\bibfnamefont{J.}~\bibnamefont{Wagner}},
  \bibinfo{journal}{Eur. Phys. J. C} \textbf{\bibinfo{volume}{81}},
  \bibinfo{pages}{300} (\bibinfo{year}{2021}{\natexlab{a}}),
  \eprint{2101.03855}.

\bibitem[{\citenamefont{Dutrieux
  et~al.}(2021{\natexlab{b}})\citenamefont{Dutrieux, Bertone, Moutarde, and
  Sznajder}}]{Dutrieux:2021ehx}
\bibinfo{author}{\bibfnamefont{H.}~\bibnamefont{Dutrieux}},
  \bibinfo{author}{\bibfnamefont{V.}~\bibnamefont{Bertone}},
  \bibinfo{author}{\bibfnamefont{H.}~\bibnamefont{Moutarde}}, \bibnamefont{and}
  \bibinfo{author}{\bibfnamefont{P.}~\bibnamefont{Sznajder}},
  \bibinfo{journal}{Eur. Phys. J. A} \textbf{\bibinfo{volume}{57}},
  \bibinfo{pages}{250} (\bibinfo{year}{2021}{\natexlab{b}}),
  \eprint{2105.09245}.

\bibitem[{\citenamefont{Paukkunen and Zurita}(2014)}]{Paukkunen:2014zia}
\bibinfo{author}{\bibfnamefont{H.}~\bibnamefont{Paukkunen}} \bibnamefont{and}
  \bibinfo{author}{\bibfnamefont{P.}~\bibnamefont{Zurita}},
  \bibinfo{journal}{JHEP} \textbf{\bibinfo{volume}{12}}, \bibinfo{pages}{100}
  (\bibinfo{year}{2014}), \eprint{1402.6623}.

\bibitem[{\citenamefont{Moutarde et~al.}(2013)\citenamefont{Moutarde, Pire,
  Sabatie, Szymanowski, and Wagner}}]{Moutarde:2013qs}
\bibinfo{author}{\bibfnamefont{H.}~\bibnamefont{Moutarde}},
  \bibinfo{author}{\bibfnamefont{B.}~\bibnamefont{Pire}},
  \bibinfo{author}{\bibfnamefont{F.}~\bibnamefont{Sabatie}},
  \bibinfo{author}{\bibfnamefont{L.}~\bibnamefont{Szymanowski}},
  \bibnamefont{and} \bibinfo{author}{\bibfnamefont{J.}~\bibnamefont{Wagner}},
  \bibinfo{journal}{Phys. Rev. D} \textbf{\bibinfo{volume}{87}},
  \bibinfo{pages}{054029} (\bibinfo{year}{2013}), \eprint{1301.3819}.

\bibitem[{\citenamefont{Braun et~al.}(2014)\citenamefont{Braun, Manashov,
  M\"uller, and Pirnay}}]{Braun:2014sta}
\bibinfo{author}{\bibfnamefont{V.~M.} \bibnamefont{Braun}},
  \bibinfo{author}{\bibfnamefont{A.~N.} \bibnamefont{Manashov}},
  \bibinfo{author}{\bibfnamefont{D.}~\bibnamefont{M\"uller}}, \bibnamefont{and}
  \bibinfo{author}{\bibfnamefont{B.~M.} \bibnamefont{Pirnay}},
  \bibinfo{journal}{Phys. Rev. D} \textbf{\bibinfo{volume}{89}},
  \bibinfo{pages}{074022} (\bibinfo{year}{2014}), \eprint{1401.7621}.

\bibitem[{\citenamefont{Guo et~al.}(2021)\citenamefont{Guo, Ji, and
  Liu}}]{Guo:2021ibg}
\bibinfo{author}{\bibfnamefont{Y.}~\bibnamefont{Guo}},
  \bibinfo{author}{\bibfnamefont{X.}~\bibnamefont{Ji}}, \bibnamefont{and}
  \bibinfo{author}{\bibfnamefont{Y.}~\bibnamefont{Liu}},
  \bibinfo{journal}{Phys. Rev. D} \textbf{\bibinfo{volume}{103}},
  \bibinfo{pages}{096010} (\bibinfo{year}{2021}), \eprint{2103.11506}.

\bibitem[{\citenamefont{An and Saghai}(2019)}]{An:2019tld}
\bibinfo{author}{\bibfnamefont{C.~S.} \bibnamefont{An}} \bibnamefont{and}
  \bibinfo{author}{\bibfnamefont{B.}~\bibnamefont{Saghai}},
  \bibinfo{journal}{Phys. Rev. D} \textbf{\bibinfo{volume}{99}},
  \bibinfo{pages}{094039} (\bibinfo{year}{2019}), \eprint{1905.05330}.

\bibitem[{\citenamefont{Lu and Schmidt}(2010)}]{Lu:2010dt}
\bibinfo{author}{\bibfnamefont{Z.}~\bibnamefont{Lu}} \bibnamefont{and}
  \bibinfo{author}{\bibfnamefont{I.}~\bibnamefont{Schmidt}},
  \bibinfo{journal}{Phys. Rev. D} \textbf{\bibinfo{volume}{82}},
  \bibinfo{pages}{094005} (\bibinfo{year}{2010}), \eprint{1008.2684}.

\bibitem[{\citenamefont{Lu and Schmidt}(2007)}]{Lu:2006kt}
\bibinfo{author}{\bibfnamefont{Z.}~\bibnamefont{Lu}} \bibnamefont{and}
  \bibinfo{author}{\bibfnamefont{I.}~\bibnamefont{Schmidt}},
  \bibinfo{journal}{Phys. Rev. D} \textbf{\bibinfo{volume}{75}},
  \bibinfo{pages}{073008} (\bibinfo{year}{2007}), \eprint{hep-ph/0611158}.

\bibitem[{\citenamefont{Bacchetta and Radici}(2011)}]{Bacchetta:2011gx}
\bibinfo{author}{\bibfnamefont{A.}~\bibnamefont{Bacchetta}} \bibnamefont{and}
  \bibinfo{author}{\bibfnamefont{M.}~\bibnamefont{Radici}},
  \bibinfo{journal}{Phys. Rev. Lett.} \textbf{\bibinfo{volume}{107}},
  \bibinfo{pages}{212001} (\bibinfo{year}{2011}), \eprint{1107.5755}.

\bibitem[{\citenamefont{Lorce and Pasquini}(2011)}]{Lorce:2011kd}
\bibinfo{author}{\bibfnamefont{C.}~\bibnamefont{Lorce}} \bibnamefont{and}
  \bibinfo{author}{\bibfnamefont{B.}~\bibnamefont{Pasquini}},
  \bibinfo{journal}{Phys. Rev. D} \textbf{\bibinfo{volume}{84}},
  \bibinfo{pages}{014015} (\bibinfo{year}{2011}), \eprint{1106.0139}.

\bibitem[{\citenamefont{Thomas}(2008)}]{Thomas:2008ga}
\bibinfo{author}{\bibfnamefont{A.~W.} \bibnamefont{Thomas}},
  \bibinfo{journal}{Phys. Rev. Lett.} \textbf{\bibinfo{volume}{101}},
  \bibinfo{pages}{102003} (\bibinfo{year}{2008}), \eprint{0803.2775}.

\bibitem[{\citenamefont{Avakian et~al.}(2008)\citenamefont{Avakian, Efremov,
  Schweitzer, and Yuan}}]{Avakian:2008dz}
\bibinfo{author}{\bibfnamefont{H.}~\bibnamefont{Avakian}},
  \bibinfo{author}{\bibfnamefont{A.~V.} \bibnamefont{Efremov}},
  \bibinfo{author}{\bibfnamefont{P.}~\bibnamefont{Schweitzer}},
  \bibnamefont{and} \bibinfo{author}{\bibfnamefont{F.}~\bibnamefont{Yuan}},
  \bibinfo{journal}{Phys. Rev. D} \textbf{\bibinfo{volume}{78}},
  \bibinfo{pages}{114024} (\bibinfo{year}{2008}), \eprint{0805.3355}.

\bibitem[{\citenamefont{She et~al.}(2009)\citenamefont{She, Zhu, and
  Ma}}]{She:2009jq}
\bibinfo{author}{\bibfnamefont{J.}~\bibnamefont{She}},
  \bibinfo{author}{\bibfnamefont{J.}~\bibnamefont{Zhu}}, \bibnamefont{and}
  \bibinfo{author}{\bibfnamefont{B.-Q.} \bibnamefont{Ma}},
  \bibinfo{journal}{Phys. Rev. D} \textbf{\bibinfo{volume}{79}},
  \bibinfo{pages}{054008} (\bibinfo{year}{2009}), \eprint{0902.3718}.

\bibitem[{\citenamefont{Filippone and Ji}(2001)}]{Filippone:2001ux}
\bibinfo{author}{\bibfnamefont{B.~W.} \bibnamefont{Filippone}}
  \bibnamefont{and} \bibinfo{author}{\bibfnamefont{X.-D.} \bibnamefont{Ji}},
  \bibinfo{journal}{Adv. Nucl. Phys.} \textbf{\bibinfo{volume}{26}},
  \bibinfo{pages}{1} (\bibinfo{year}{2001}), \eprint{hep-ph/0101224}.

\bibitem[{\citenamefont{Kuhn et~al.}(2009)\citenamefont{Kuhn, Chen, and
  Leader}}]{Kuhn:2008sy}
\bibinfo{author}{\bibfnamefont{S.~E.} \bibnamefont{Kuhn}},
  \bibinfo{author}{\bibfnamefont{J.~P.} \bibnamefont{Chen}}, \bibnamefont{and}
  \bibinfo{author}{\bibfnamefont{E.}~\bibnamefont{Leader}},
  \bibinfo{journal}{Prog. Part. Nucl. Phys.} \textbf{\bibinfo{volume}{63}},
  \bibinfo{pages}{1} (\bibinfo{year}{2009}), \eprint{0812.3535}.

\bibitem[{\citenamefont{Liu and Lorc\'e}(2016)}]{Liu:2015xha}
\bibinfo{author}{\bibfnamefont{K.-F.} \bibnamefont{Liu}} \bibnamefont{and}
  \bibinfo{author}{\bibfnamefont{C.}~\bibnamefont{Lorc\'e}},
  \bibinfo{journal}{Eur. Phys. J. A} \textbf{\bibinfo{volume}{52}},
  \bibinfo{pages}{160} (\bibinfo{year}{2016}), \eprint{1508.00911}.

\bibitem[{\citenamefont{Kumeri\v{c}ki et~al.}(2014)\citenamefont{Kumeri\v{c}ki,
  M\"uller, and Murray}}]{Kumericki:2013br}
\bibinfo{author}{\bibfnamefont{K.}~\bibnamefont{Kumeri\v{c}ki}},
  \bibinfo{author}{\bibfnamefont{D.}~\bibnamefont{M\"uller}}, \bibnamefont{and}
  \bibinfo{author}{\bibfnamefont{M.}~\bibnamefont{Murray}},
  \bibinfo{journal}{Phys. Part. Nucl.} \textbf{\bibinfo{volume}{45}},
  \bibinfo{pages}{723} (\bibinfo{year}{2014}), \eprint{1301.1230}.

\bibitem[{\citenamefont{Shiells et~al.}(2022)\citenamefont{Shiells, Guo, and
  Ji}}]{Shiells:2021xqo}
\bibinfo{author}{\bibfnamefont{K.}~\bibnamefont{Shiells}},
  \bibinfo{author}{\bibfnamefont{Y.}~\bibnamefont{Guo}}, \bibnamefont{and}
  \bibinfo{author}{\bibfnamefont{X.}~\bibnamefont{Ji}}, \bibinfo{journal}{JHEP}
  \textbf{\bibinfo{volume}{08}}, \bibinfo{pages}{048} (\bibinfo{year}{2022}),
  \eprint{2112.15144}.

\bibitem[{\citenamefont{Kriesten et~al.}(2020)\citenamefont{Kriesten, Liuti,
  Calero-Diaz, Keller, Meyer, Goldstein, and Osvaldo
  Gonzalez-Hernandez}}]{Kriesten:2019jep}
\bibinfo{author}{\bibfnamefont{B.}~\bibnamefont{Kriesten}},
  \bibinfo{author}{\bibfnamefont{S.}~\bibnamefont{Liuti}},
  \bibinfo{author}{\bibfnamefont{L.}~\bibnamefont{Calero-Diaz}},
  \bibinfo{author}{\bibfnamefont{D.}~\bibnamefont{Keller}},
  \bibinfo{author}{\bibfnamefont{A.}~\bibnamefont{Meyer}},
  \bibinfo{author}{\bibfnamefont{G.~R.} \bibnamefont{Goldstein}},
  \bibnamefont{and} \bibinfo{author}{\bibfnamefont{J.}~\bibnamefont{Osvaldo
  Gonzalez-Hernandez}}, \bibinfo{journal}{Phys. Rev. D}
  \textbf{\bibinfo{volume}{101}}, \bibinfo{pages}{054021}
  (\bibinfo{year}{2020}), \eprint{1903.05742}.

\bibitem[{\citenamefont{Kriesten
  et~al.}(2022{\natexlab{b}})\citenamefont{Kriesten, Liuti, and
  Meyer}}]{Kriesten:2020apm}
\bibinfo{author}{\bibfnamefont{B.}~\bibnamefont{Kriesten}},
  \bibinfo{author}{\bibfnamefont{S.}~\bibnamefont{Liuti}}, \bibnamefont{and}
  \bibinfo{author}{\bibfnamefont{A.}~\bibnamefont{Meyer}},
  \bibinfo{journal}{Phys. Lett. B} \textbf{\bibinfo{volume}{829}},
  \bibinfo{pages}{137051} (\bibinfo{year}{2022}{\natexlab{b}}),
  \eprint{2011.04484}.

\bibitem[{\citenamefont{Kriesten and Liuti}(2022)}]{Kriesten:2020wcx}
\bibinfo{author}{\bibfnamefont{B.}~\bibnamefont{Kriesten}} \bibnamefont{and}
  \bibinfo{author}{\bibfnamefont{S.}~\bibnamefont{Liuti}},
  \bibinfo{journal}{Phys. Rev. D} \textbf{\bibinfo{volume}{105}},
  \bibinfo{pages}{016015} (\bibinfo{year}{2022}), \eprint{2004.08890}.

\bibitem[{\citenamefont{He et~al.}(2022)\citenamefont{He, Ji, Melnitchouk,
  Thomas, and Wang}}]{He:2022leb}
\bibinfo{author}{\bibfnamefont{F.}~\bibnamefont{He}},
  \bibinfo{author}{\bibfnamefont{C.-R.} \bibnamefont{Ji}},
  \bibinfo{author}{\bibfnamefont{W.}~\bibnamefont{Melnitchouk}},
  \bibinfo{author}{\bibfnamefont{A.~W.} \bibnamefont{Thomas}},
  \bibnamefont{and} \bibinfo{author}{\bibfnamefont{P.}~\bibnamefont{Wang}},
  \bibinfo{journal}{Phys. Rev. D} \textbf{\bibinfo{volume}{106}},
  \bibinfo{pages}{054006} (\bibinfo{year}{2022}), \eprint{2202.00266}.

\bibitem[{\citenamefont{Xu et~al.}(2021)\citenamefont{Xu, Mondal, Lan, Zhao,
  Li, and Vary}}]{Xu:2021wwj}
\bibinfo{author}{\bibfnamefont{S.}~\bibnamefont{Xu}},
  \bibinfo{author}{\bibfnamefont{C.}~\bibnamefont{Mondal}},
  \bibinfo{author}{\bibfnamefont{J.}~\bibnamefont{Lan}},
  \bibinfo{author}{\bibfnamefont{X.}~\bibnamefont{Zhao}},
  \bibinfo{author}{\bibfnamefont{Y.}~\bibnamefont{Li}}, \bibnamefont{and}
  \bibinfo{author}{\bibfnamefont{J.~P.} \bibnamefont{Vary}}
  (\bibinfo{collaboration}{BLFQ}), \bibinfo{journal}{Phys. Rev. D}
  \textbf{\bibinfo{volume}{104}}, \bibinfo{pages}{094036}
  (\bibinfo{year}{2021}), \eprint{2108.03909}.

\bibitem[{\citenamefont{Liu et~al.}(2022)\citenamefont{Liu, Xu, Mondal, Zhao,
  and Vary}}]{Liu:2022fvl}
\bibinfo{author}{\bibfnamefont{Y.}~\bibnamefont{Liu}},
  \bibinfo{author}{\bibfnamefont{S.}~\bibnamefont{Xu}},
  \bibinfo{author}{\bibfnamefont{C.}~\bibnamefont{Mondal}},
  \bibinfo{author}{\bibfnamefont{X.}~\bibnamefont{Zhao}}, \bibnamefont{and}
  \bibinfo{author}{\bibfnamefont{J.~P.} \bibnamefont{Vary}}
  (\bibinfo{collaboration}{BLFQ}), \bibinfo{journal}{Phys. Rev. D}
  \textbf{\bibinfo{volume}{105}}, \bibinfo{pages}{094018}
  (\bibinfo{year}{2022}), \eprint{2202.00985}.

\bibitem[{\citenamefont{Lin et~al.}(2018)}]{Lin:2017snn}
\bibinfo{author}{\bibfnamefont{H.-W.} \bibnamefont{Lin}} \bibnamefont{et~al.},
  \bibinfo{journal}{Prog. Part. Nucl. Phys.} \textbf{\bibinfo{volume}{100}},
  \bibinfo{pages}{107} (\bibinfo{year}{2018}), \eprint{1711.07916}.

\bibitem[{\citenamefont{Constantinou et~al.}(2021)}]{Lin:2020rut}
\bibinfo{author}{\bibfnamefont{M.}~\bibnamefont{Constantinou}}
  \bibnamefont{et~al.}, \bibinfo{journal}{Prog. Part. Nucl. Phys.}
  \textbf{\bibinfo{volume}{121}}, \bibinfo{pages}{103908}
  (\bibinfo{year}{2021}), \eprint{2006.08636}.

\bibitem[{\citenamefont{Lin}(2022)}]{Lin:2021brq}
\bibinfo{author}{\bibfnamefont{H.-W.} \bibnamefont{Lin}},
  \bibinfo{journal}{Phys. Lett. B} \textbf{\bibinfo{volume}{824}},
  \bibinfo{pages}{136821} (\bibinfo{year}{2022}), \eprint{2112.07519}.

\bibitem[{\citenamefont{Lin}(2021)}]{Lin:2020rxa}
\bibinfo{author}{\bibfnamefont{H.-W.} \bibnamefont{Lin}},
  \bibinfo{journal}{Phys. Rev. Lett.} \textbf{\bibinfo{volume}{127}},
  \bibinfo{pages}{182001} (\bibinfo{year}{2021}), \eprint{2008.12474}.

\bibitem[{\citenamefont{Alexandrou et~al.}(2020)\citenamefont{Alexandrou,
  Cichy, Constantinou, Hadjiyiannakou, Jansen, Scapellato, and
  Steffens}}]{Alexandrou:2020zbe}
\bibinfo{author}{\bibfnamefont{C.}~\bibnamefont{Alexandrou}},
  \bibinfo{author}{\bibfnamefont{K.}~\bibnamefont{Cichy}},
  \bibinfo{author}{\bibfnamefont{M.}~\bibnamefont{Constantinou}},
  \bibinfo{author}{\bibfnamefont{K.}~\bibnamefont{Hadjiyiannakou}},
  \bibinfo{author}{\bibfnamefont{K.}~\bibnamefont{Jansen}},
  \bibinfo{author}{\bibfnamefont{A.}~\bibnamefont{Scapellato}},
  \bibnamefont{and} \bibinfo{author}{\bibfnamefont{F.}~\bibnamefont{Steffens}},
  \bibinfo{journal}{Phys. Rev. Lett.} \textbf{\bibinfo{volume}{125}},
  \bibinfo{pages}{262001} (\bibinfo{year}{2020}), \eprint{2008.10573}.

\bibitem[{\citenamefont{Alexandrou et~al.}(2022)\citenamefont{Alexandrou,
  Cichy, Constantinou, Hadjiyiannakou, Jansen, Scapellato, and
  Steffens}}]{Alexandrou:2021bbo}
\bibinfo{author}{\bibfnamefont{C.}~\bibnamefont{Alexandrou}},
  \bibinfo{author}{\bibfnamefont{K.}~\bibnamefont{Cichy}},
  \bibinfo{author}{\bibfnamefont{M.}~\bibnamefont{Constantinou}},
  \bibinfo{author}{\bibfnamefont{K.}~\bibnamefont{Hadjiyiannakou}},
  \bibinfo{author}{\bibfnamefont{K.}~\bibnamefont{Jansen}},
  \bibinfo{author}{\bibfnamefont{A.}~\bibnamefont{Scapellato}},
  \bibnamefont{and} \bibinfo{author}{\bibfnamefont{F.}~\bibnamefont{Steffens}},
  \bibinfo{journal}{Phys. Rev. D} \textbf{\bibinfo{volume}{105}},
  \bibinfo{pages}{034501} (\bibinfo{year}{2022}), \eprint{2108.10789}.

\bibitem[{\citenamefont{Dodson et~al.}(2022)\citenamefont{Dodson, Bhattacharya,
  Cichy, Constantinou, Metz, Scapellato, and Steffens}}]{Bhattacharya:2021oyr}
\bibinfo{author}{\bibfnamefont{J.}~\bibnamefont{Dodson}},
  \bibinfo{author}{\bibfnamefont{S.}~\bibnamefont{Bhattacharya}},
  \bibinfo{author}{\bibfnamefont{K.}~\bibnamefont{Cichy}},
  \bibinfo{author}{\bibfnamefont{M.}~\bibnamefont{Constantinou}},
  \bibinfo{author}{\bibfnamefont{A.}~\bibnamefont{Metz}},
  \bibinfo{author}{\bibfnamefont{A.}~\bibnamefont{Scapellato}},
  \bibnamefont{and} \bibinfo{author}{\bibfnamefont{F.}~\bibnamefont{Steffens}},
  \bibinfo{journal}{PoS} \textbf{\bibinfo{volume}{LATTICE2021}},
  \bibinfo{pages}{054} (\bibinfo{year}{2022}), \eprint{2112.05538}.

\bibitem[{\citenamefont{Hannaford-Gunn
  et~al.}(2022)\citenamefont{Hannaford-Gunn, Can, Horsley, Nakamura, Perlt,
  Rakow, St\"uben, Schierholz, Young, and Zanotti}}]{CSSMQCDSFUKQCD:2021lkf}
\bibinfo{author}{\bibfnamefont{A.}~\bibnamefont{Hannaford-Gunn}},
  \bibinfo{author}{\bibfnamefont{K.~U.} \bibnamefont{Can}},
  \bibinfo{author}{\bibfnamefont{R.}~\bibnamefont{Horsley}},
  \bibinfo{author}{\bibfnamefont{Y.}~\bibnamefont{Nakamura}},
  \bibinfo{author}{\bibfnamefont{H.}~\bibnamefont{Perlt}},
  \bibinfo{author}{\bibfnamefont{P.~E.~L.} \bibnamefont{Rakow}},
  \bibinfo{author}{\bibfnamefont{H.}~\bibnamefont{St\"uben}},
  \bibinfo{author}{\bibfnamefont{G.}~\bibnamefont{Schierholz}},
  \bibinfo{author}{\bibfnamefont{R.~D.} \bibnamefont{Young}}, \bibnamefont{and}
  \bibinfo{author}{\bibfnamefont{J.~M.} \bibnamefont{Zanotti}}
  (\bibinfo{collaboration}{CSSM/QCDSF/UKQCD}), \bibinfo{journal}{Phys. Rev. D}
  \textbf{\bibinfo{volume}{105}}, \bibinfo{pages}{014502}
  (\bibinfo{year}{2022}), \eprint{2110.11532}.

\bibitem[{\citenamefont{Engelhardt}(2017)}]{Engelhardt:2017miy}
\bibinfo{author}{\bibfnamefont{M.}~\bibnamefont{Engelhardt}},
  \bibinfo{journal}{Phys. Rev. D} \textbf{\bibinfo{volume}{95}},
  \bibinfo{pages}{094505} (\bibinfo{year}{2017}), \eprint{1701.01536}.

\bibitem[{\citenamefont{Engelhardt et~al.}(2020)\citenamefont{Engelhardt,
  Green, Hasan, Krieg, Meinel, Negele, Pochinsky, and
  Syritsyn}}]{Engelhardt:2020qtg}
\bibinfo{author}{\bibfnamefont{M.}~\bibnamefont{Engelhardt}},
  \bibinfo{author}{\bibfnamefont{J.~R.} \bibnamefont{Green}},
  \bibinfo{author}{\bibfnamefont{N.}~\bibnamefont{Hasan}},
  \bibinfo{author}{\bibfnamefont{S.}~\bibnamefont{Krieg}},
  \bibinfo{author}{\bibfnamefont{S.}~\bibnamefont{Meinel}},
  \bibinfo{author}{\bibfnamefont{J.}~\bibnamefont{Negele}},
  \bibinfo{author}{\bibfnamefont{A.}~\bibnamefont{Pochinsky}},
  \bibnamefont{and} \bibinfo{author}{\bibfnamefont{S.}~\bibnamefont{Syritsyn}},
  \bibinfo{journal}{Phys. Rev. D} \textbf{\bibinfo{volume}{102}},
  \bibinfo{pages}{074505} (\bibinfo{year}{2020}), \eprint{2008.03660}.

\bibitem[{\citenamefont{Benali et~al.}(2020)}]{Benali:2020vma}
\bibinfo{author}{\bibfnamefont{M.}~\bibnamefont{Benali}} \bibnamefont{et~al.},
  \bibinfo{journal}{Nature Phys.} \textbf{\bibinfo{volume}{16}},
  \bibinfo{pages}{191} (\bibinfo{year}{2020}), \eprint{2109.02076}.

\bibitem[{\citenamefont{Ellinghaus et~al.}(2006)\citenamefont{Ellinghaus,
  Nowak, Vinnikov, and Ye}}]{Ellinghaus:2005uc}
\bibinfo{author}{\bibfnamefont{F.}~\bibnamefont{Ellinghaus}},
  \bibinfo{author}{\bibfnamefont{W.~D.} \bibnamefont{Nowak}},
  \bibinfo{author}{\bibfnamefont{A.~V.} \bibnamefont{Vinnikov}},
  \bibnamefont{and} \bibinfo{author}{\bibfnamefont{Z.}~\bibnamefont{Ye}},
  \bibinfo{journal}{Eur. Phys. J. C} \textbf{\bibinfo{volume}{46}},
  \bibinfo{pages}{729} (\bibinfo{year}{2006}), \eprint{hep-ph/0506264}.

\end{thebibliography}


\end{document}